\documentclass[aps,showpacs,twocolumn,longbibliography]{revtex4-1}

\usepackage[english]{babel}
\usepackage[utf8]{inputenc}
\usepackage{amsmath}
\usepackage{amssymb}
\usepackage[caption=false]{subfig}
\usepackage{amssymb}
\usepackage{epsfig}
\usepackage{graphicx}
\usepackage{amsmath}
\usepackage{array,color}
\usepackage{natbib}

\usepackage[usenames,dvipsnames]{xcolor}
\definecolor{forestgreen}{rgb}{0.11,0.54,0.15}
\definecolor{purple}{rgb}{0.62,0.10,0.96}
\definecolor{dockerblue}{rgb}{0.11,0.56,0.98}
\definecolor{freeblue}{rgb}{0.25,0.41,0.88}

\usepackage[pdftex,plainpages=false,colorlinks=true,linkcolor=Red, citecolor=blue, urlcolor=blue]{hyperref}

\newcommand{\opj}{\hat{\jmath}}
\newcommand{\opo}{\hat{\mathcal{O}}}

\newcommand{\vk}{\vec{k}}

\begin{document}

\title{Maximum entropy analytic continuation for frequency-dependent transport coefficients with non-positive spectral weight}
\author{A. Reymbaut$^1$, A.-M. Gagnon$^1$, D. Bergeron$^1$, and A.-M. S. Tremblay$^{1,2}$}
\affiliation{
$^1$D\'{e}partement de physique and Institut quantique, Universit\'{e} de Sherbrooke, Sherbrooke, Qu\'{e}bec, Canada J1K 2R1 \\
$^2$Canadian Institute for Advanced Research, Toronto, Ontario, Canada, M5G 1Z8
}
\date{\today}
\begin{abstract}
The computation of transport coefficients, even in linear response, is a major challenge for theoretical methods that rely on analytic continuation of correlations functions obtained numerically in Matsubara space. While maximum entropy methods can be used for certain correlation functions, this is not possible in general, important examples being the Seebeck, Hall, Nernst and Reggi-Leduc coefficients. Indeed, positivity of the spectral weight on the positive real-frequency axis is not guaranteed in these cases. The spectral weight can even be complex in the presence of broken time-reversal symmetry. Various workarounds, such as the neglect of vertex corrections or the study of the infinite frequency or Kelvin limits have been proposed. Here, we show that one can define auxiliary response functions that allow to extract the desired real-frequency susceptibilities from maximum entropy methods in the most general multiorbital cases with no particular symmetry. As a benchmark case, we study the longitudinal thermoelectric response and corresponding Onsager coefficient in the single-band two-dimensional Hubbard model treated with dynamical mean-field theory (DMFT) and continuous-time quantum Monte Carlo (CTQMC). We thereby extend to transport coefficients the maximum entropy analytic continuation with auxiliary functions (MaxEntAux method), developed for the study of the superconducting pairing dynamics of correlated materials.
\end{abstract}
\pacs{71.27.+a, 72.10.-d, 72.10.Bg, 72.15.Jf, 72.20.Pa}
\maketitle



\paragraph{Introduction.} 

Transport properties are of interest for both fundamental and applied purposes. For example, while thermoelectric power tells us about the nature of charge carriers, materials with large thermoelectric power could lead to various applications, including efficient conversion of heat loss into useful electricity \cite{Terasaki:1997, Bentien:2007, Tomczak:2010, Sun:2011, Arsenault:2013, Ozaeta:2014, Kolenda:2016}. Unfortunately, computing transport coefficients from numerical results is no simple task. Usually, one starts by computing the corresponding response functions in Matsubara frequency using the Kubo formula. For quantum Monte-Carlo data in particular, the most direct way to extract the real-frequency dependent response functions is then to perform maximum entropy analytic continuations (MEACs) \cite{Jarrell:1996, Sandvik:1998}. However, as we explain in more details below, MEAC is not always trivial since it requires that the spectral weight of response functions is real and positive, which is not necessarily the case in general.  

Many approaches have been investigated to circumvent this major problem for the Seebeck coefficient \cite{Beni:1974, Chaikin:1976, Palsson:1998, Koshibae:2000, Oudovenko:2002, Kontani:2003, Shastry:2006, Shastry:2009, Chakraborty:2010, Xu:2011}, the Hall coefficient \cite{Shastry:1993, Assaad:1995, Kumar:2003, Kumar:2004, Haerter:2008} and the Nernst coefficient \cite{Xu:2013} for instance, but all of them are either approximations or analytic methods that are exact only in a certain frequency limit~\footnote{However, see the recent work \protect\cite{OtsukiShinaokaMaxEnt:2017}}. The most common approach consists in neglecting vertex corrections, in which case it is possible to compute transport coefficients directly from the single-particle spectral weight. This is not possible when vertex corrections are included, which seems to be a necessary step in understanding the record-breaking thermopowers of FeSb$_2$ and FeAs$_2$ \cite{Tomczak:2010} for instance. In that case, a more versatile approach, that would extend to all transport quantities, still remains to be developed. This is the problem that we address here by generalizing the MaxEntAux method developed for fermionic response functions~\cite{Reymbaut:2015_MaxEnt,Reymbaut:2016_Cuprates_1} to the case of bosonic response functions with non-positive  spectral weights~\cite{GagnonMSc:2016}. We first describe this method for the most general multiorbital system with no particular symmetry before presenting a benchmark case for the uniform longitudinal thermoelectric response (or uniform Seebeck response) of a single-band two-dimensional Hubbard model treated with dynamical mean-field theory (DMFT)~\cite{Georges:1996} and continuous-time quantum Monte Carlo (CTQMC)~\cite{Gull:2011}. 
The DC limit is also compared to a low-temperature approximation~\cite{Randeria:1992}, detailed in the supplemental material Ref.~\footnote{See supplemental material at [] for the choice of convention for the Onsager coefficients (identical to that of Ref.~\protect\cite{Mahan:2000}). Explanations behind the low-temperature approximation Eq.~\protect\eqref{Eq_L_Randeria}~\protect\cite{Randeria:1992} are given. We also present results at $U=14$, $T = 1/10$ and $T=1$ including the response functions' spectral weights along with a more thorough comparison with Ref.~\protect\cite{Xu:2011} and the approximation Eq.~\protect\eqref{Eq_L_Randeria}. It is also shown how the results are affected by the choice of  $\lambda$ in Eqs.~\protect\eqref{Eq_mixed_operator-agd} and \protect\eqref{Eq_mixed_operator_b}.  We end with new expressions~\protect\cite{GagnonMSc:2016} for the bubble part of the uniform susceptibilities that are convergent upon summation over internal Matsubara frequencies~\protect\cite{Bergeron:2011}. The convergence of the corresponding expressions given in Ref.~\cite{Paul:2003} is discussed. In addition, we include examples of data files used for obtaining the first figure of this paper with OmegaMaxEnt~\protect\cite{Bergeron:2015}. The supplemental material also contains a set of data files enabling the user to reproduce the T=1 and lambda=0.3 results of the parent paper using the OmegaMaxEnt code~\protect\cite{Bergeron:2015}.}. Some of the issues associated with convergence of Matsubara frequency sums are also discussed there~\cite{Note2}.


\paragraph{Bosonic response functions without positive spectral weight} Let us define the correlation function between general bosonic operators $\opo_{\vk\gamma}$ and $\opo_{\vk\delta}$ by
\begin{equation}
\chi_{\gamma\delta}(\vec{k},\tau) = - \left\langle \hat{\mathcal{T}}_{\tau}\, \opo_{\vk\gamma}(\tau) \, \opo_{\vk\delta}^\dagger(0) \right\rangle_{\hat{\mathcal{H}}} \, ,
\label{Eq_correlation_function}
\end{equation}
where $\vec{k}$ is wave vector, the average is taken with respect to the grand-canonical ensemble for Hamiltonian $\hat{\mathcal{H}}$ (with eigenvectors $\vert m \rangle$ and eigenenergies $H_m$) and $\hat{\mathcal{T}_{\tau}}$ is the bosonic imaginary-time ordering operator. In the case of current operators, that we consider in this paper, $\gamma$ and $\delta$ stand for the electrical ($E$) or thermal ($T$) nature of the current operators and also for spatial direction ($x$, $y$ or $z$), band, and spin indices. The discrete Fourier transform to bosonic Matsubara frequency space, $\chi_{\gamma\delta}(\vec{k},i\omega_n) = \int_0^\beta \! \mathrm{d}\tau\, e^{i\omega_n\tau} \chi_{\gamma\delta}(\vec{k},\tau)$, is related to the spectral weight $\chi^{\prime\prime}_{\gamma\delta}(\vec{k},\omega)$ through the relation
\begin{equation}
\chi_{\gamma\delta}(\vk,i\omega_n) = \int_0^\beta \! \mathrm{d}\tau\, e^{i\omega_n\tau} \chi_{\gamma\delta}(\vk,\tau) = \int\!\frac{\mathrm{d}\omega}{\pi}\,\frac{\chi^{\prime\prime}_{\gamma\delta}(\vk,\omega)}{\omega-i\omega_n} \, .
\label{Eq_Spectral_representation_Matsubara}
\end{equation}
This spectral weight is essential, for example, to obtain the finite-frequency behavior of the Hall conductivity (related to the Faraday effect) that obeys the analog of a $f$-sum rule \cite{Spielman:1994, Drew_Coleman:1997}. 

The non-triviality of the MEAC for a general response function arises because this method requires that 
\begin{equation}
\forall \omega, \; \omega\, \chi^{\prime\prime}_{\gamma\delta}(\vec{k},\omega) \geq 0 \quad \text{and} \quad \chi^{\prime\prime}_{\gamma\delta}(\vec{k},\omega) \in \mathbb{R}
\label{Condition_MEAC}
\end{equation}
for bosonic data. However, the Lehmann representation
\begin{eqnarray}
 & & \frac{\pi}{\mathcal{Z}}\sum_{mm'} e^{-\beta H_m}\left[ e^{\beta \omega} - 1 \right] \langle m | \opo_{\vk\gamma} |m'\rangle \langle m' | \opo_{\vk\delta}^\dagger|m\rangle \nonumber \\
 & & \times  \delta(\omega - (H_m - H_{m'}))
\label{Eq_Lehmann}
\end{eqnarray}
for $\chi^{\prime\prime}_{\gamma\delta}(\vec{k},\omega)$, where $\mathcal{Z}$ is the partition function of the system, tells us that this condition is obviously satisfied if $\delta \equiv \gamma$ since 
\begin{equation}
\langle m | \opo_{\vk\gamma} |m'\rangle \langle m' | \opo^\dagger_{\vk\gamma}|m\rangle = \vert \langle m | \opo_{\vk\gamma} |m'\rangle \vert^2 \geq 0 \, .
\end{equation}
and $e^{\beta \omega} - 1$ is positive for $\omega > 0$ and negative for $\omega < 0$. This special case corresponds to purely electric or purely thermal longitudinal responses, hence transverse response functions in the presence of a symmetry-breaking magnetic field or even, more simply, the thermoelectric response, cannot be obtained with the standard approach.



\paragraph{MaxEntAux Method for the General Case.} Consider the general correlation function Eq.~\eqref{Eq_correlation_function}. Generalizing the approach of Ref.~\cite{Reymbaut:2015_MaxEnt}, we define the mixed operator
\begin{equation}
\hat{\mathcal{A}}_{\vk\gamma\delta,\lambda} = \opo_{\vk\gamma} + \lambda\,\opo^\dagger_{-\vk\delta} \, .
\label{Eq_mixed_operator-agd}
\end{equation}
With the usual Matsubara imaginary-time evolution $\opo_{\vk\gamma}(\tau)=e^{H\tau}\opo_{\vk\gamma} e^{-H\tau}$ and $\opo^\dagger_{\vk\gamma}(\tau)=e^{H\tau}\opo_{\vk\gamma}^\dagger e^{-H\tau}$ we define the auxiliary susceptibility 
\begin{equation}
\chi^{aux\,1}_{\gamma\delta,\lambda}(\vk,\tau) = -\left\langle \hat{\mathcal{T}}_{\tau}\, \hat{\mathcal{A}}_{\vk\gamma\delta,\lambda}(\tau)\, \hat{\mathcal{A}}^\dagger_{\vk\gamma\delta,\lambda}(0) \right\rangle_{\hat{\mathcal{H}}} \, ,
\label{Eq_Total_Green_Function}
\end{equation}
where $\lambda$ is an arbitrary real constant that, in principle, ensures conversion of units between two possibly different current operators in $\hat{\mathcal{A}}_{\vk\gamma\delta,\lambda}$. 
The Lehmann representation allows one to check that the auxiliary susceptibility Eq.~\eqref{Eq_Total_Green_Function} satisfies the condition Eq.~\eqref{Condition_MEAC} and can thus be analytically continued using standard maximum entropy methods. With hermitian operators, we have  $\opo_{-\vk\gamma}^\dagger=\opo_{\vk\gamma}$ so that one finds
\begin{eqnarray}
\chi^{aux\,1}_{\gamma\delta,\lambda}(\vk,\tau) & = & \chi_{\gamma\gamma} (\vk,\tau) + \lambda^2 \chi_{\delta\delta} (\vk,\tau) \nonumber \\  
 & & + \lambda\,\chi_{\gamma\delta}(\vk,\tau) + \lambda\,[\chi_{\gamma\delta}(\vk,\tau)]^* \, . 
\label{Gaux1}
\end{eqnarray}
Using the spectral representation Eq.~\eqref{Eq_Spectral_representation_Matsubara} for the susceptibility, Eq.~\eqref{Gaux1} can then be obtained in Matsubara frequencies:
\begin{eqnarray}
\chi^{aux\,1}_{\gamma\delta,\lambda}(\vk,i\omega_n)  &=&  \chi_{\gamma\gamma} (\vk,i\omega_n) + \lambda^2\chi_{\delta\delta} (\vk,i\omega_n) \nonumber \\ 
  & & + \lambda\,\chi_{\gamma\delta}(\vk,i\omega_n) +\lambda\,[\chi_{\gamma\delta}(\vk,-i\omega_n)]^* 
 \label{Eq_to_reduce}\\
&=&  \chi_{\gamma\gamma} (\vk,i\omega_n) + \lambda^2\chi_{\delta\delta} (\vk,i\omega_n) \nonumber \\
 & &  + \lambda \!\int\!\frac{\mathrm{d}\omega}{\pi}\,\frac{\chi^{\prime\prime}_{\gamma\delta}(\vk,\omega) + \chi^{\prime\prime\,*}_{\gamma\delta}(\vk,\omega)}{\omega-i\omega_n} \, .
\label{Gaux1Matsubara}
\end{eqnarray}
To find the missing information, we define, by analogy with Eq.~\eqref{Eq_mixed_operator-agd}, a second mixed operator
\begin{equation}
\hat{\mathcal{B}}_{\vk\gamma\delta,\lambda^\prime} = \opo_{\vk\gamma} + i\lambda^\prime\,\opo^\dagger_{-\vk\delta} \, ,
\label{Eq_mixed_operator_b}
\end{equation}
and a corresponding second auxiliary susceptibility satisfying the condition Eq.~\eqref{Condition_MEAC}, 
\begin{equation}
\chi^{aux\,2}_{\gamma\delta,\lambda^\prime}(\vec{k},\tau) = -\left\langle \hat{\mathcal{T}}_{\tau}\, \hat{\mathcal{B}}_{\vk\gamma\delta,\lambda^\prime}(\tau)\, \hat{\mathcal{B}}^{\dagger}_{\vk\gamma\delta,\lambda^\prime}(0) \right\rangle_{\hat{\mathcal{H}}} \, .
\label{Eq_Total_Green_Function-2}
\end{equation}
Then, by analogy with Eq.~\eqref{Gaux1Matsubara}, we find
\begin{eqnarray}
\chi^{aux\,2}_{\gamma\delta,\lambda^\prime}(\vk,i\omega_n) & = & \chi_{\gamma\gamma} (\vk,i\omega_n) + \lambda^{\prime 2}\chi_{\delta\delta} (\vk,i\omega_n) \nonumber \\
 & &  -i\lambda^\prime\!\int\!\frac{\mathrm{d}\omega}{\pi}\,\frac{\chi^{\prime\prime}_{\gamma\delta}(\vk,\omega) - \chi^{\prime\prime\,*}_{\gamma\delta}(\vk,\omega)}{\omega-i\omega_n} \, .
\label{Gaux2Matsubara}
\end{eqnarray}

The MEAC of the two auxiliary susceptibilities gives the corresponding real spectral weights satisfying the condition Eq.~\eqref{Condition_MEAC}, which are formally given by
\begin{widetext}
\begin{eqnarray}
\chi^{\prime\prime\, aux\,1}_{\gamma\delta,\lambda}(\vec{k},\omega) & = & \chi^{\prime\prime}_{\gamma\gamma}(\vk,\omega)+\lambda^2\chi^{\prime\prime}_{\delta\delta}(\vk,\omega) + \lambda\left[\chi^{\prime\prime}_{\gamma\delta}(\vk,\omega)+\chi^{\prime\prime\,*}_{\gamma\delta}(\vk,\omega)\right] \, , \\
\chi^{\prime\prime\, aux\,2}_{\gamma\delta,\lambda^\prime}(\vk,\omega) & = & \chi^{\prime\prime}_{\gamma\gamma}(\vk,\omega)+\lambda^{\prime 2}\chi^{\prime\prime}_{\delta\delta}(\vk,\omega)  -i\lambda^\prime\left[\chi^{\prime\prime}_{\gamma\delta}(\vk,\omega)-\chi^{\prime\prime\,*}_{\gamma\delta}(\vk,\omega)\right]\, .
\end{eqnarray}
The spectral weights of the form $\chi^{\prime\prime}_{\gamma\gamma}(\vk,\omega)$ also satisfy the condition Eq.~\eqref{Condition_MEAC}. From the above, one can easily extract the needed complex-valued spectral weight $\chi^{\prime\prime}_{\gamma\delta}(\vec{k},\omega)$. Taking  $\lambda=\lambda^\prime$ for simplicity, we obtain 
\begin{eqnarray}
\chi^{\prime\prime}_{\gamma\delta}(\vk,\omega) & = & \frac{1}{2\lambda}  \left\{ \chi^{\prime\prime\,aux\,1}_{\gamma\delta,\lambda}(\vk,\omega) + i\, \chi^{\prime\prime\, aux\,2}_{\gamma\delta,\lambda}(\vk,\omega)  -(1+i)[\chi^{\prime\prime}_{\gamma\gamma}(\vk,\omega) + \lambda^2\chi^{\prime\prime}_{\delta\delta}(\vk,\omega) ] \right\}\, .
\end{eqnarray}
\end{widetext}
In the case of current operators, this is valid for any kind of response function in the multi-orbital case, with or without inversion or time-reversal symmetry. Taking $\lambda\ne\lambda^\prime$ would allow more flexibility to balance various contributions.


\paragraph{Practical Example for the Uniform Seebeck Response.} The study of uniform longitudinal thermoelectricity relies only on the three response functions for electrical and thermal currents $\chi_{E_x E_x}$, $\chi_{E_x T_x}$ and $\chi_{T_x T_x}$, more generally defined as:
\begin{equation}
\chi_{\gamma \delta}(\tau) = - \left\langle \hat{\mathcal{T}}_{\tau}\, \opj_{\gamma}(\tau) \, \opj_{\delta}(0) \right\rangle_{\hat{\mathcal{H}}} \, ,
\label{Chi_alpha}
\end{equation}
where we explicitly wrote the current operators $\opj_{\gamma}$, standing for the electrical and thermal currents at $\vec{k}=\vec{0}$ (with $\opj^\dagger_{\vec{k}\gamma} = \opj_{-\vec{k}\gamma}$ in general). As mentioned above, while $\chi_{E_x E_x}$ and $\chi_{T_x T_x}$ can be analytically continued through standard maximum entropy methods, $\chi_{E_x T_x}$ cannot be. For this practical example, we focus on a system with time-reversal symmetry. In that case, one can show that any spectral weight $\chi^{\prime\prime}_{\gamma\delta}(\omega)$ is real and odd in real frequency when the two operators involved have the same signature under time-reversal. The spectral representation Eq.~\eqref{Eq_Spectral_representation_Matsubara} then implies that $\chi_{\gamma\delta}(i\omega_n)$ is real and even in Matsubara frequency, thus Eq.~\eqref{Eq_to_reduce} can be rewritten as 
\begin{equation}
\chi_{E_x T_x}(i\omega_n)=\frac{1}{2\lambda}(\chi^{aux}_\lambda(i\omega_n)
- \chi_{E_x E_x}(i\omega_n) - \lambda^2\chi_{T_x T_x}(i\omega_n)  )\, .
\label{Eq_Trick_iwn_uniform}
\end{equation}

Defining the frequency-dependent Onsager coefficients through
\begin{equation}
\mathrm{Re} \,\mathcal{L}_{\gamma\delta}(\omega) =  \lim\limits_{\vec{k} \to \vec{0}}\, \left[ \frac{\chi^{\prime\prime}_{\gamma\delta}(\vec{k},\omega)}{\omega}\right]\, .
\label{Eq_Onsager_coefficients}
\end{equation}
the bosonic Matsubara frequency result Eq.~\eqref{Eq_Trick_iwn_uniform} immediately translates to 
\begin{equation}
\mathcal{L}_{E_x T_x}(\omega) = \frac{1}{2\lambda} \left[ \mathcal{L}^{aux}_\lambda(\omega) - \mathcal{L}_{E_x E_x}(\omega) - \lambda^2 \mathcal{L}_{T_x T_x}(\omega) \right] \, .
\label{Eq_Trick_uniform_L}
\end{equation}

\paragraph{Model and method} We consider the two-dimensional square lattice described by the single-band Hubbard model
\begin{equation}
\hat{\mathcal{H}} = -t\sum_{\langle i, j\rangle \; \sigma} \left(\hat{c}^\dagger_{i\sigma} \, \hat{c}_{j\sigma}+ h.c\right) + U \sum_{i} \hat{n}_{i\uparrow} \, \hat{n}_{i\downarrow} - \mu \sum_i \hat{n}_i \, ,
\label{Hubbard_model}
\end{equation}
with $t$ the nearest-neighbor hopping, $U$ the on-site Coulomb repulsion and $\mu$ the chemical potential. We take $t\equiv 1$ as our energy unit, lattice spacing as our distance unit, and otherwise take natural units ($\hbar \equiv 1$, $k_B \equiv 1$ and electrical charge $e=1$) . 

When vertex corrections are included, the direct MEAC of $\chi_{\gamma\delta}(i\omega_n)$ is mandatory. But to benchmark our approach, we consider a case where vertex corrections are neglected so that the transport coefficients can be computed directly from the single particle spectral weight using
\begin{eqnarray}
\chi^{\prime\prime}_{\nu}(\omega) & = & \pi T \sum_{\sigma} \int_{-4t}^{4t} \! \mathrm{d}\varepsilon \int \! \mathrm{d}\omega'\, \mathcal{T}(\varepsilon)\, \mathcal{A}(\varepsilon, \omega')\, \mathcal{A}(\varepsilon, \omega' + \omega) \nonumber \\
& & \times \left[ f(\omega') - f(\omega' + \omega) \right] \left( \omega' + \frac{\omega}{2} \right)^\nu\, ,
\label{Analytic_expression_chi}
\end{eqnarray}
where $f$ is the Fermi-Dirac distribution and $\mathcal{A}(\vec{k},\omega)$ is the spectral function containing the non-interacting square-lattice dispersion $\varepsilon_{\vec{k}}$ and normalized so that $\int \! \mathrm{d}\omega\, \mathcal{A}(\vec{k},\omega) = 1$. In this notation, $\chi^{\prime\prime}_{\nu=0}(\omega) = \chi^{\prime\prime}_{E_xE_x}(\omega)$, $\chi^{\prime\prime}_{\nu=1}(\omega) = \chi^{\prime\prime}_{E_xT_x}(\omega)$, and $\chi^{\prime\prime}_{\nu=2}(\omega) = \chi^{\prime\prime}_{T_xT_x}(\omega)$ \cite{Xu:2011}. Here, the usual integral over wave-vectors has been replaced by an integral over the band energies $\varepsilon$ weighted by the longitudinal transport function \cite{LFA:2013} 
\begin{equation}
\mathcal{T}(\varepsilon) = \sum_{\vec{k}} \left( \frac{\partial \varepsilon_{\vec{k}}}{\partial k_x} \right)^2 \delta(\varepsilon - \varepsilon_{\vec{k}}) = -\frac{1}{2}\int_{-4t}^{\varepsilon} \! z N_0(z) \, \mathrm{d}z
\end{equation}  
containing the non-interacting density of states $N_0$, normalized so that $\int \! N_0(z)\, \mathrm{d}z = 1$. 

Our test is performed as follows. Given the value of $\chi_\nu^{\prime\prime}(\omega)$ in Eq.~\eqref{Analytic_expression_chi}, we obtain Matsubara susceptibilities $\chi_\nu(i\omega_n)$ from the spectral representation Eq.~\eqref{Eq_Spectral_representation_Matsubara}. We then add a Gaussian noise with a relative error of $10^{-3}$ on each Matsubara frequency in order to mimic quantum Monte-Carlo statistical error. Finally, we use the maximum entropy code OmegaMaxEnt \cite{Bergeron:2015} to extract the analytically continued spectral weights $\chi_\nu^{\prime\prime}(\omega)$, that can then be compared with the starting values. While $\chi_{E_x E_x}(i\omega_n)$ and $\chi_{T_x T_x}(i\omega_n)$ are analytically continued directly, the MaxEntAux formula Eq.~\eqref{Eq_Trick_iwn_uniform} is used to analytically continue $\chi_{E_x T_x}(i\omega_n)$. 

Any physical single-particle spectral weight $\mathcal{A}(\vec{k},\omega)$ could have been used for the test, but to be as realistic as possible, we used one obtained from an actual calculation. We take $U=14$, temperature $T=1$, and set the chemical potential so that filling is $n=0.85$, a case studied in Ref.~\cite{Xu:2011}. It is at high-temperature that analytic continuation is most difficult and that the sign change in the frequency-dependent thermoelectric spectral weight is largest~\cite{Xu:2011}. A more thorough comparison and results at a lower temperature $T = 1/10$ are presented in the supplemental material Ref.~\cite{Note2}. We use dynamical mean-field theory (DMFT)~\cite{Georges:1996}  solved with continuous-time quantum Monte Carlo (CTQMC)~\cite{Gull:2011} to simulate our system and compute the local fermionic Matsubara-frequency self-energy $\Sigma(ip_n)$ which is analytically continued to give us our starting $\mathcal{A}(\vec{k},\omega)$ and corresponding $\chi_\nu^{\prime\prime}(\omega)$.

\begin{figure}[h!]
	\begin{center}
		\includegraphics[width=0.45\textwidth]{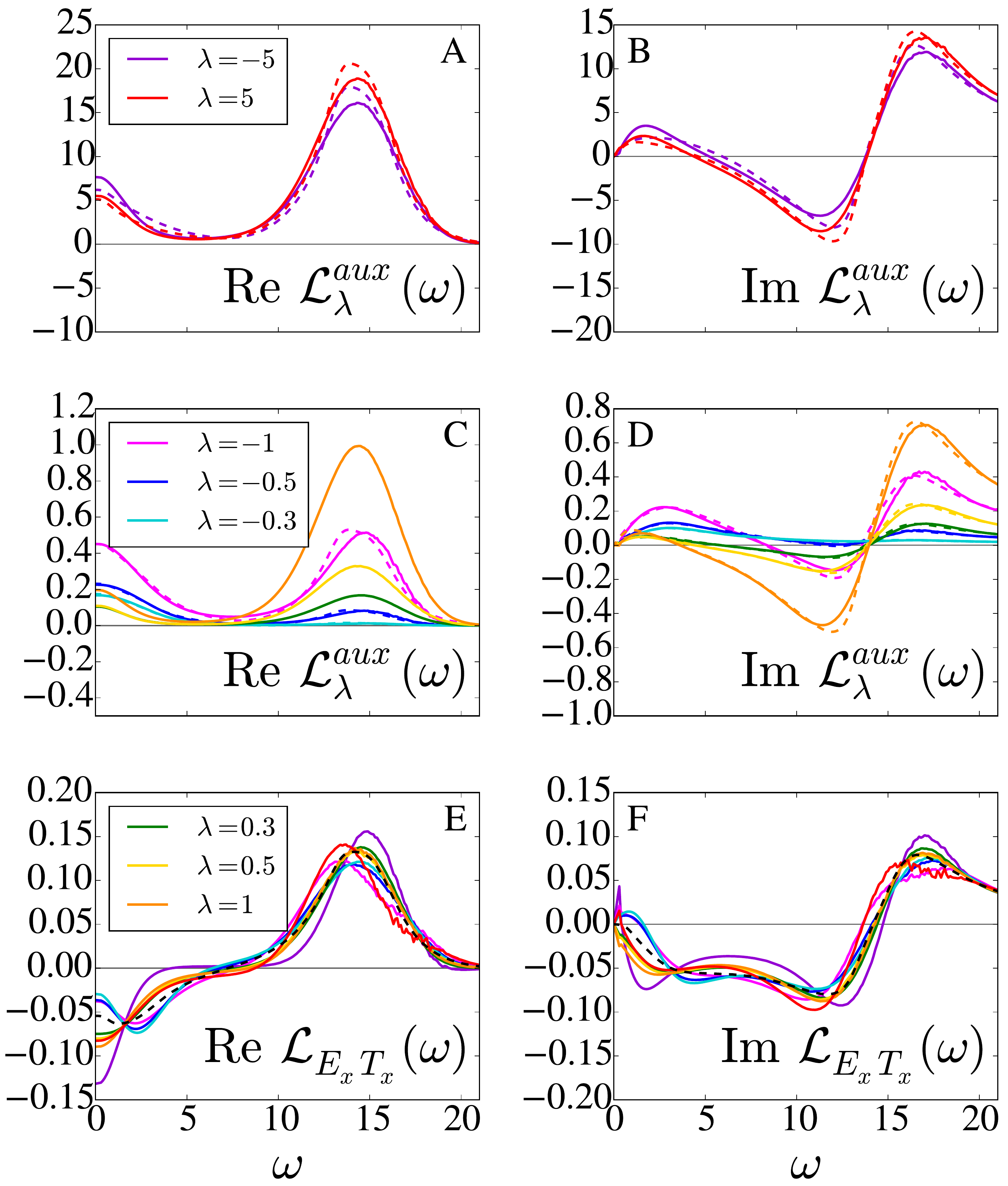}
		\caption{Comparison for $T=1$, $U=14$ between the initial auxiliary and thermoelectric frequency-dependent Onsager coefficients Eq.~\eqref{Eq_Onsager_coefficients} calculated with Eq.~\eqref{Analytic_expression_chi} (dashed lines) and those obtained for different values of $\lambda$ from Eq.~\eqref{Eq_Trick_uniform_L} after analytic continuation of the Matsubara data with an added relative noise of $10^{-3}$ using the MaxEntAux method (solid lines). The black dashed lines in the lowest panels represent the expected thermoelectric frequency-dependent Onsager coefficient obtained from Eq.~\eqref{Analytic_expression_chi} using Eq.~\eqref{Eq_Onsager_coefficients}. The color coding for the values of $\lambda $ is spread over all panels.}
		\label{Fig_L}
	\end{center}
\end{figure}

\paragraph{Results.} Figs.~\ref{Fig_L}~A-D compare the initial (dashed lines) real-frequency auxiliary Onsager coefficients $\mathcal{L}^{aux}_{\lambda}(\omega)$ obtained from $\chi^{\prime\prime}_{\nu}(\omega)$ Eqs.~\eqref{Eq_Onsager_coefficients} and \eqref{Analytic_expression_chi} to those obtained after MEAC (solid lines) for different values of $\lambda$. These auxiliary functions are $\lambda$-dependent by definition. The physical thermoelectric response $\mathcal{L}_{E_xT_x}(\omega)$ in Fig.~\ref{Fig_L}~E-F is extracted from the MaxEntAux method, i.e. from the analytic continuation of the right-hand side of Eq.~\eqref{Eq_Trick_uniform_L}. The result should be $\lambda$-independent. The agreement with the benchmark black dashed line is generally good, except for $\lambda=-5$ and at low frequency, as we discuss further below. The real part of $\mathcal{L}_{E_xT_x}(\omega)$ has a large region of negative spectral weight that is reproduced. The imaginary part of $\mathcal{L}_{E_xT_x}(\omega)$ is obtained from Kramers-Kronig.

\begin{figure}[h!]
	\begin{center}
		\includegraphics[width=0.40\textwidth]{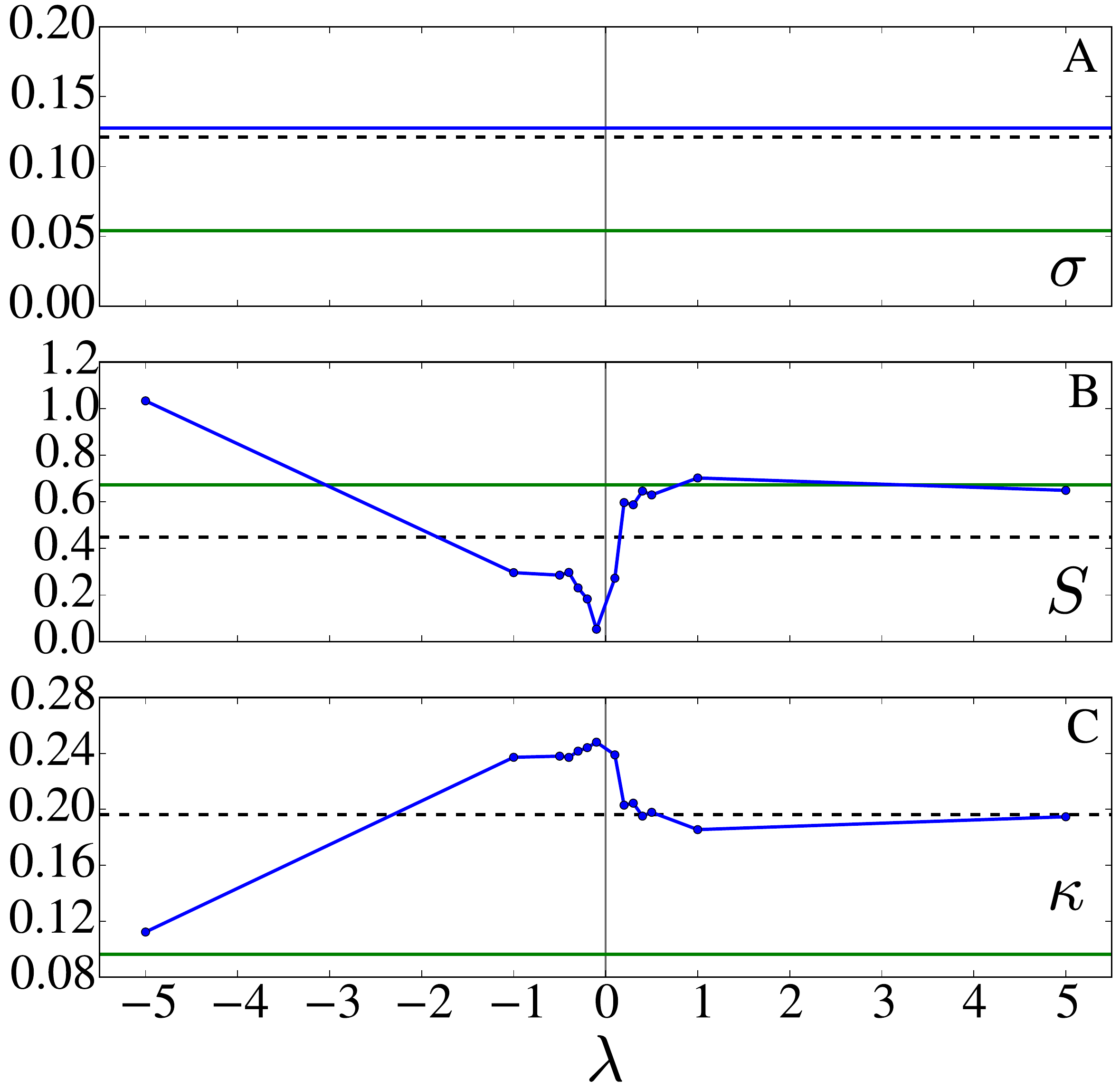}
		\caption{Electrical conductivity, Seebeck coefficient, and thermal conductivity for $T=1,U=14$ from Eqs.~\eqref{Eq_sigma},~\eqref{Eq_S}, and \eqref{Eq_kappa} respectively. Comparison between the benchmark values (black dashed lines) and those of the MaxEntAux method applied to the Matsubara frequency results with an added relative noise of $10^{-3}$ (blue solid line in panel A and blue points in panels B and C). The blue solid lines in panels B and C are guides to the eye for the $\lambda$-dependent results. The green lines show the results of the low-temperature approximation Eq.~\eqref{Eq_L_Randeria}.
			}
		\label{Fig_sigma_S}
	\end{center}
\end{figure}

To assess the choice of $\lambda$, consider Fig.~\ref{Fig_sigma_S} that shows the electrical conductivity $\sigma$, the Seebeck coefficient $S$ and the thermal conductivity $\kappa$, defined by
\begin{equation}
\sigma = \frac{1}{T}\, \mathcal{L}_{E_xE_x}\, , \label{Eq_sigma}
\end{equation}
\begin{equation}
S = -\frac{1}{T}\, \frac{\mathcal{L}_{E_xT_x}}{\mathcal{L}_{E_xE_x}} \, , \label{Eq_S}
\end{equation}
and
\begin{equation}
\kappa = \frac{1}{T^2}\, \left[ \mathcal{L}_{T_xT_x} - \frac{\mathcal{L}_{E_xT_x}^2}{\mathcal{L}_{E_xE_x}} \right] , \label{Eq_kappa}
\end{equation}
where the convention~\cite{Mahan:2000, Xu:2011} chosen for the Onsager coefficients $\mathcal{L}_{\gamma\delta}=\lim_{\omega \to 0}\,\mathcal{L}_{\gamma\delta}(\omega)$ is presented in the supplemental material Ref.~\cite{Note2}. Since we use natural units the Seebeck coefficient is retrieved by multiplying our value by $k_B/e = 86.3~\mu\mathrm{V}/\mathrm{K}$. 

The blue solid line in Fig.~\ref{Fig_sigma_S}A shows that the electrical conductivity obtained after MEAC is within 10 \% of the benchmark Eq.~\eqref{Analytic_expression_chi} (black dashed line).  Although with perfect analytic continuation the $\lambda$-dependence of $\mathcal{L}_{E_xT_x}$ in the MaxEntAux expression Eq.~\eqref{Eq_Trick_uniform_L} would drop out, in practice it does not, as shown for $S$ and $\kappa$ in Figs.~\ref{Fig_sigma_S}~B-C by the blue solid lines linking calculated blue points. Clearly the $1/\lambda$ prefactor in Eq.~\eqref{Eq_Trick_uniform_L} suggests that a small value for $|\lambda|$ enhances errors. Also, since $\textrm{Re}\,\mathcal{L}_{E_xT_x}(\omega)$ is positive for most of the positive frequency range, large negative values of $\lambda$ mean that $\textrm{Re}\,\mathcal{L}_{T_xT_x}(\omega)$ is most important in the calculation of $\textrm{Re}\,\mathcal{L}_{E_xT_x}(\omega)$. But this is the most difficult transport coefficient to analytically continue because of its sensitive dependency on low frequencies. Therefore, one should only consider that the best result is obtained for the range of values $\lambda \geq 0.5$ where the $\lambda$-dependency is weak in Figs.~\ref{Fig_sigma_S}~B-C. 
Finally, the comparison with the low-temperature green-line approximation \cite{Randeria:1992,Chen:2016} $\mathcal{L}_{\gamma\delta} \simeq (\beta/\pi)^2\,\mathcal{L}_{\gamma\delta}(\tau=\beta/2)$ or
\begin{equation}
\mathcal{L}_{\gamma\delta} \simeq  \frac{\beta^2}{\pi^2}\int_{-\infty}^{+\infty}\! \mathrm{d} \omega\, \frac{\chi^{\prime\prime}_{\gamma\delta}(\vec{0},\omega)}{2\sinh\left( \frac{\beta\omega}{2} \right)} 
\label{Eq_L_Randeria}
\end{equation}
shows that, for the case studied here, the MaxEntAux method gives more accurate values than Refs.~\cite{Randeria:1992,Chen:2016} for all transport coefficients~\footnote{Notice that this approximation is not valid in the renormalized classical regime since, in this regime, the convergence of the integral in Eq.~\protect\eqref{Eq_L_Randeria} is controlled by the susceptibility instead of the $\sinh$ factor.}. The supplemental material Ref.~\protect\cite{Note2} discusses lower temperatures where Eq.~\protect\eqref{Eq_L_Randeria} becomes more competitive.

\paragraph{Conclusion.} 

The above results open the way to the systematic exploration of frequency-dependent transport properties obtained from numerical calculations in Matsubara space with or without vertex corrections. While benchmarks were shown for the uniform  Seebeck response $\chi^{\prime\prime}_{E_x T_x}(\omega)$, even less trivial problems can be investigated. For instance, the Hall $\chi^{\prime\prime}_{E_x E_y}(\vec{k},\omega)$ and Nernst $\chi^{\prime\prime}_{E_x T_y}(\vec{k},\omega)$ response functions are of interest in cases where time-reversal symmetry is broken by a magnetic field.

\paragraph{Acknowledgments.} We are indebted to S.~Verret, D.~Sénéchal, J.~Gukelberger, R.~Nourafkan, B.~S.~Shastry and Wenhu Xu for fruitful discussions. This work has been supported by the Natural Sciences and Engineering Research Council of Canada (NSERC) under grant RGPIN-2014-04584, and by the Research Chair in the Theory of Quantum Materials (A.-M.S. T.). Simulations were performed on computers provided by Canadian Foundation for Innovation, the Minist\`ere de l’\'Education des Loisirs et du Sport (Qu\'ebec), Calcul Qu\'ebec, and Compute Canada.

\appendix
\newpage
\setcounter{section}{0}

\section{Summary of supplemental material}

In Sec.~\ref{Sec:Convention}, we detail the choice of convention for the Onsager coefficients, identical to the one in Ref.~[\onlinecite{Mahan:2000}].  
In Sec.~\ref{Sec:Low_T_approx}, we explain the low-temperature approximation taken from Ref.~[\onlinecite{Randeria:1992}]. We recapitulate our procedure for benchmarks and present in Sec.~\ref{Sec:Susceptibilities} additional results at $T = 1/10$ and $T=1$ for the spectral weights of relevant susceptibilities. We continue with more benchmark tests on the Onsager coefficients and a more thorough comparison with the low-temperature approximation from Ref.~[\onlinecite{Randeria:1992}] in Sec.~\ref{Sec:More_tests}. Combinations of Onsager coefficients are related to the transport coefficients. We benchmark those in Sec.~\ref{Sec:Transport-Coeff}. Finally, in Sec.~\ref{Sec:ConvergenceMatsubara}, we give new expressions, taken from Ref.~[\onlinecite{GagnonMSc:2016}], for the bubble part of the uniform susceptibilities that are convergent upon summation over internal Matsubara frequencies. The convergence of the corresponding expressions given in Ref.~[\onlinecite{Paul:2003}] is discussed. Note that the independence of the results on the value of $\lambda$ is discussed in Sec.~\ref{Sec:Susceptibilities} and Sec.~\ref{Sec:More_tests}.

In this supplemental material, we do not consider the case where a magnetic field is applied. In other words, time-reversal invariance is assumed.

\begin{widetext}
\section{Convention for the Onsager Coefficients.} \label{Sec:Convention} The relationship between the Onsager coefficients 
\begin{equation}
 \mathcal{L}_{\gamma\delta} = \lim\limits_{\omega \to 0}\, \mathrm{Re}\,\mathcal{L}_{\gamma\delta}(\omega) = \lim\limits_{\omega \to 0}\, \lim\limits_{\vec{k} \to \vec{0}}\, \left[ \frac{\chi^{\prime\prime}_{\gamma\delta}(\vec{k},\omega)}{\omega}\right]
\label{Eq_Onsager_coefficients_sup}
\end{equation}
and the relevant transport quantities always depends on  the choice made for the form of the phenomenological equations describing how electrical and heat currents are induced by gradients of electrical potential and temperature in the uniform and static limits. For instance, in the case of the longitudinal thermoelectric response (or Seebeck response), we chose the same equations as Ref.~[\onlinecite{Mahan:2000}]:
\begin{eqnarray}
\vec{\jmath}_{E_x} 
& = & \mathcal{L}_{E_xE_x} \left[ - \frac{1}{T}\, \vec{\nabla}_{x}(\mu-eV) \right] + \mathcal{L}_{E_xT_x} \left[ \vec{\nabla}_{x} \frac{1}{T} \right] \, , \\ 
\vec{\jmath}_{T_x} 
& = & \mathcal{L}_{T_xE_x} \left[ - \frac{1}{T}\, \vec{\nabla}_{x}(\mu-eV) \right] + \mathcal{L}_{T_xT_x} \left[ \vec{\nabla}_{x} \frac{1}{T} \right] \, ,
\end{eqnarray}
where $\mu$ is the chemical potential, $e$ is the electrical charge, and $V$ is the electrical potential. These equations, identical to the ones chosen in Ref.~[\onlinecite{Xu:2011}], ease the comparison of our data with the results of this reference. Besides, they lead, in the absence of vertex corrections, to the following convenient general expressions for the response functions:
\begin{eqnarray}
\chi^{\prime\prime}_{\nu}(\omega) & = & \pi T \sum_{\sigma} \int \! \mathrm{d}\varepsilon \int \! \mathrm{d}\omega'\, \mathcal{T}(\varepsilon)\, \mathcal{A}(\varepsilon, \omega')\, \mathcal{A}(\varepsilon, \omega' + \omega) \nonumber \\
 & & \times \left[ f(\omega') - f(\omega' + \omega) \right] \left( \omega' + \frac{\omega}{2} \right)^\nu\, ,
 \label{Analytic_expression_chi_sup}
\end{eqnarray}
with $\nu = 0$, 1, 2 $\equiv E_xE_x$, $E_xT_x$, $T_xT_x$, respectively. This choice of convention leads to
\begin{eqnarray}
\sigma & = & \frac{1}{T}\, \mathcal{L}_{E_xE_x}\, , \label{Eq_sigma_sup} \\
S & = & -\frac{1}{T}\, \frac{\mathcal{L}_{E_xT_x}}{\mathcal{L}_{E_xE_x}} \, , \label{Eq_S_sup} \\
\kappa & = & \frac{1}{T^2}\, \left[ \mathcal{L}_{T_xT_x} - \frac{\mathcal{L}_{E_xT_x}^2}{\mathcal{L}_{E_xE_x}} \right] , \label{Eq_kappa_sup}
\end{eqnarray}
for the electrical conductivity $\sigma$, the Seebeck coefficient $S$ and the thermal conductivity $\kappa$.

\section{Origins of the Low-Temperature Approximation for $\mathcal{L}_{\gamma\delta}$.}\label{Sec:Low_T_approx} To avoid analytic continuation all together, it has been proposed to use  $\mathcal{L}_{\gamma\delta} \simeq (\beta/\pi)^2\,\mathcal{L}_{\gamma\delta}(\tau=\beta/2)$ as an approximation for the zero frequency transport coefficients at low temperature~[\onlinecite{Randeria:1992}]. We have compared our results with that approximation, rewritten as:
\begin{equation}
\mathcal{L}_{\gamma\delta} \simeq  \frac{\beta^2}{\pi^2}\int_{-\infty}^{+\infty}\! \mathrm{d} \omega\, \frac{\chi^{\prime\prime}_{\gamma\delta}(\vec{0},\omega)}{2\sinh\left( \frac{\beta\omega}{2} \right)}\; ,
\label{Eq_L_Randeria_sup}
\end{equation}
whose derivation we give below. 

Considering the right-hand term in Eq.~\eqref{Eq_L_Randeria_sup}, we assume that the spectral weight of the response function is linear in frequency on the whole frequency range relevant for the integral, usually controlled at low temperature by the $\sinh$ factor. We thus have $\chi^{\prime\prime}_{\gamma\delta}(\vec{0},\omega) \simeq \omega \, \mathcal{L}_{\gamma\delta}$. Since
\begin{equation}
\int_{-\infty}^{+\infty}\! \mathrm{d} \omega\, \frac{\chi^{\prime\prime}_{\gamma\delta}(\vec{0},\omega)}{2\sinh\left( \frac{\beta\omega}{2} \right)} = \int_{-\infty}^{+\infty}\! \mathrm{d} \omega\, \frac{\omega}{2\sinh\left( \frac{\beta\omega}{2} \right)} \, \frac{\chi^{\prime\prime}_{\gamma\delta}(\vec{0},\omega)}{\omega}\, ,
\end{equation}
we can approximate
\begin{equation}
\int_{-\infty}^{+\infty}\! \mathrm{d} \omega\, \frac{\chi^{\prime\prime}_{\gamma\delta}(\vec{0},\omega)}{2\sinh\left( \frac{\beta\omega}{2} \right)} \simeq \frac{2\mathcal{L}_{\gamma\delta}}{\beta^2} \int_{-\infty}^{+\infty}\! \mathrm{d} x\, \frac{x}{\sinh(x)} \, .
\end{equation}
Finally, the value of the integral
\begin{equation}
\int_{-\infty}^{+\infty}\! \mathrm{d} x\, \frac{x}{\sinh(x)} = \frac{\pi^2}{2}
\end{equation}
gives the expected result Eq.~\eqref{Eq_L_Randeria_sup}. When the convergence of the integral is controlled by the susceptibility instead of the $\sinh$ factor, as in the renormalized classical regime, a different expression has to be devised.

\begin{figure}[b!]
	\begin{minipage}{0.47\linewidth}
		\subfloat[]{\includegraphics[width=0.91\textwidth]{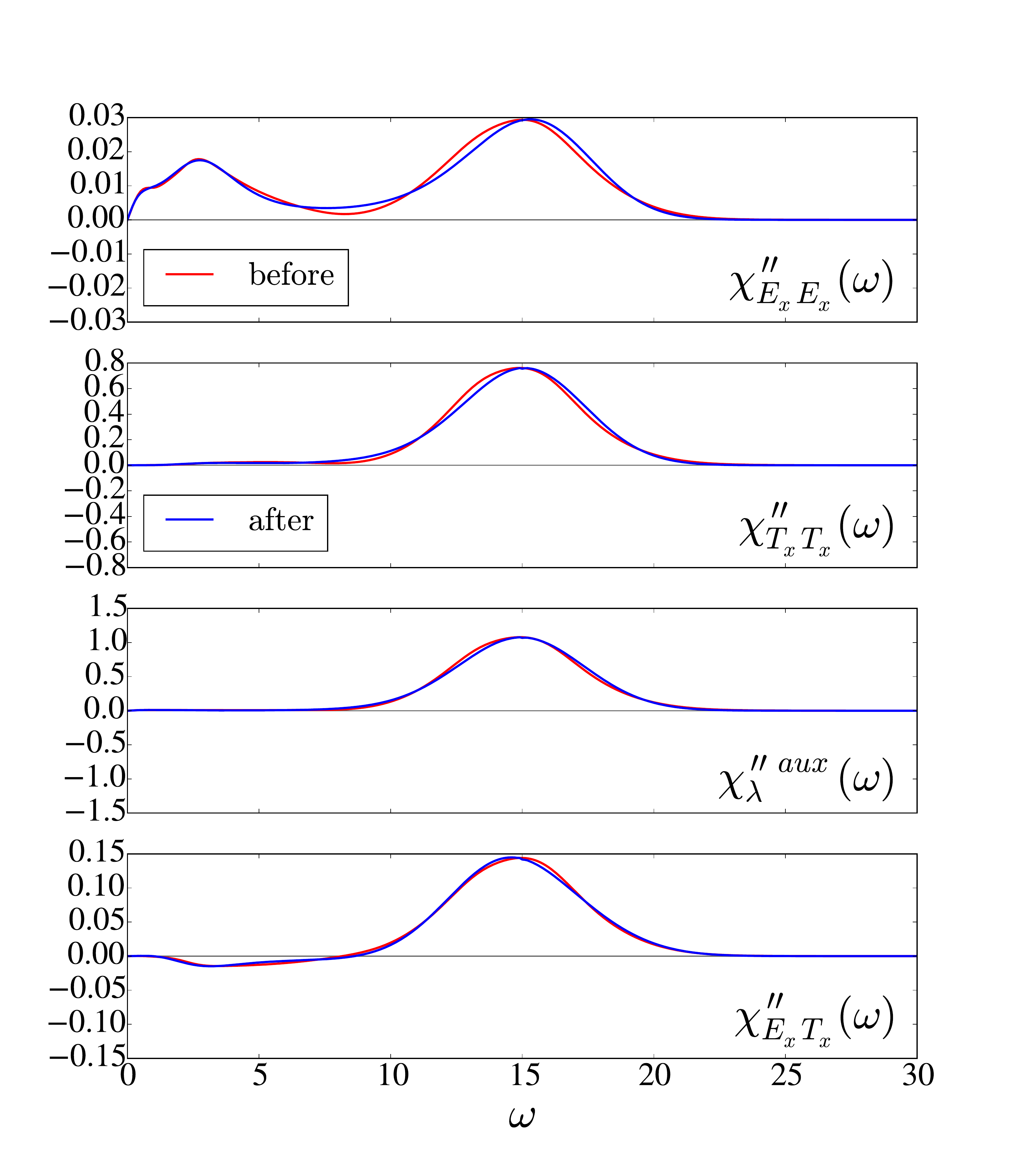} \label{Fig_Im_Chi_beta10_sup}} 
	\end{minipage}
	\begin{minipage}{0.47\linewidth}
		\subfloat[]{\includegraphics[width=0.90\textwidth]{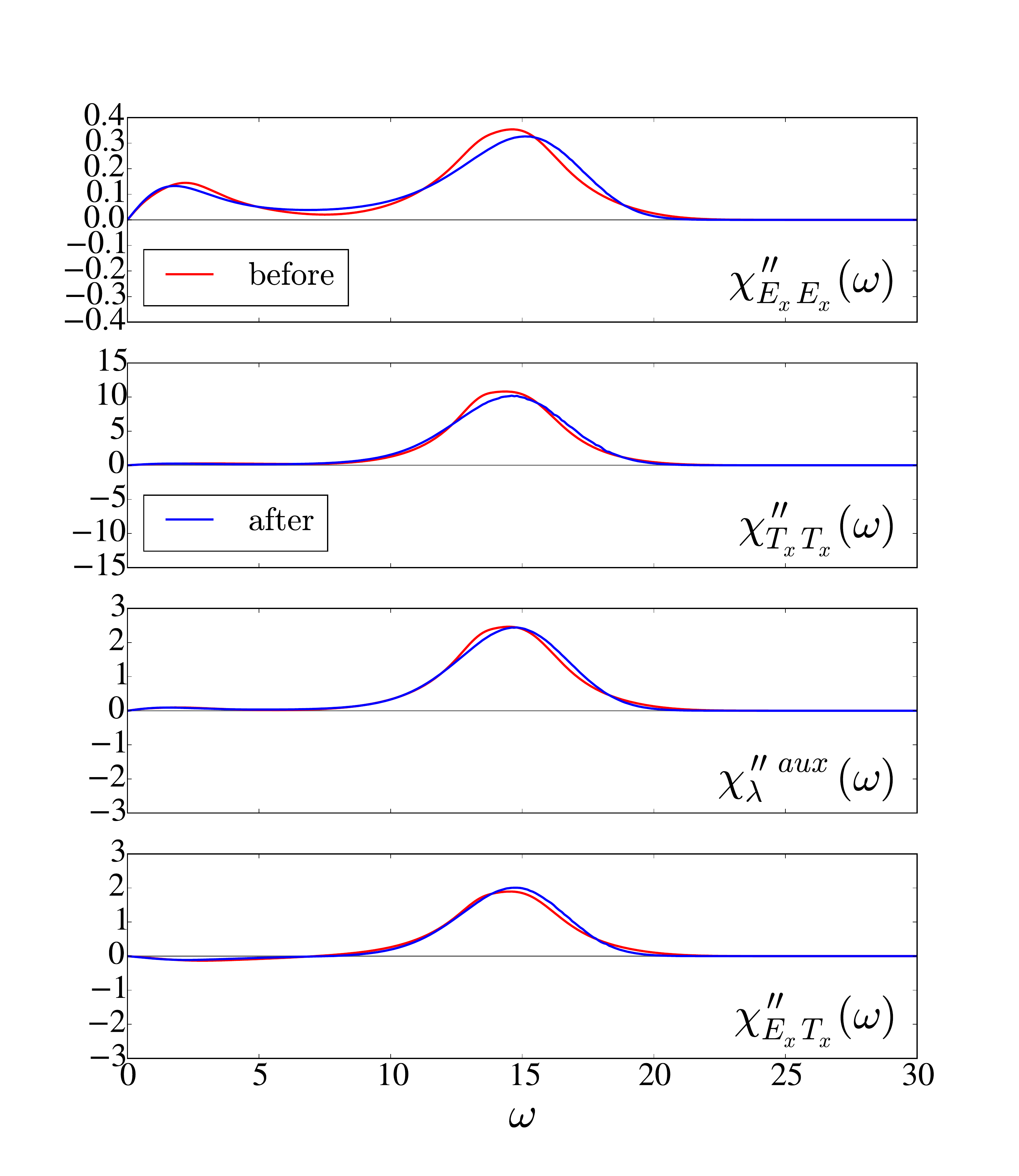} \label{Fig_Im_Chi_beta1_sup}}
	\end{minipage}
	\centering
	\caption{Comparison between the initial real-frequency response functions' spectral weights (in red) and the ones obtained after direct MEACs of the Matsubara-frequency data and the MaxEntAux method (in blue)  Eq.~\eqref{Eq_Trick_uniform_sup}. (a) Results for $\lambda=1$ at $T=1/10$. (b) Results for $\lambda=0.3$ at $T=1$.}
	\label{Fig_Im_Chi_sup}
\end{figure}

\section{Analytically Continued Spectral Weights of the Relevant Susceptibilities.}\label{Sec:Susceptibilities} Let us first recapitulate our procedure for analytic continuation before we present additional results. In the case of the Seebeck response, the MaxEntAux method consists in the use of the auxiliary susceptibility 
\begin{equation}
\chi^{aux}_\lambda(i\omega_n)
= \chi_{E_x E_x}(i\omega_n) + \lambda^2\chi_{T_x T_x}(i\omega_n) + 2\lambda\,\chi_{E_x T_x}(i\omega_n)
\label{Eq_Trick_iwn_uniform_sup}
\end{equation}
to enable the maximum entropy analytic continuation (MEAC) of the response function
\begin{equation}
\chi_{E_x T_x}^{\prime\prime}(\omega) = \frac{1}{2\lambda} \left[ \chi^{\prime \prime\, aux}_\lambda(\omega) - \chi^{\prime\prime}_{E_x E_x}(\vec{k},\omega) - \lambda^2 \chi^{\prime\prime}_{T_x T_x}(\omega) \right] \, ,
\label{Eq_Trick_uniform_sup}
\end{equation}
which possibly has a non-positive value of  the spectral weight $\chi^{\prime\prime}_{\gamma\delta}(\vec{k},\omega)/\omega$, since, according to the Lehmann representation
\begin{equation}
\chi^{\prime\prime}_{\gamma\delta}(\vec{k},\omega) = \frac{\pi}{\mathcal{Z}}\sum_{mm'} e^{-\beta H_m}\left[ e^{\beta \omega} - 1 \right] \langle m | \opo_{\gamma} |m'\rangle \langle m' | \opo_{\delta}|m\rangle\, \delta(\omega - (H_m - H_{m'}))\, ,
\label{Eq_Lehmann_sup}
\end{equation}
the sign of the matrix elements may be negative or positive and they do not simplify when $\gamma \ne \delta$. Fig.~\ref{Fig_Im_Chi_beta10_sup} and Fig.~\ref{Fig_Im_Chi_beta1_sup} compare the initial real-frequency spectral weights Eq.~\eqref{Analytic_expression_chi_sup} to those obtained after direct MEACs of the Matsubara-frequency data
\begin{equation}
\chi_{\gamma\delta}(i\omega_n) = \int_0^\beta \! \mathrm{d}\tau\, e^{i\omega_n\tau} \chi_{\gamma\delta}(\tau) = \int\!\frac{\mathrm{d}\omega}{\pi}\,\frac{\chi^{\prime\prime}_{\gamma\delta}(\omega)}{\omega-i\omega_n}\, ,
\label{Eq_Spectral_representation_Matsubara_sup}
\end{equation}
to which we add a Gaussian relative error of $10^{-3}$ on each Matsubara frequency in order to mimic quantum Monte-Carlo statistical error. For the thermoelectric response, 
the MaxEntAux method Eq.~\eqref{Eq_Trick_uniform_sup} for $\lambda=1$ at $U=14$, $T=1/10$ and for $\lambda=0.3$ at $T=1$, respectively, is used. The MaxEntAux method proves very efficient at capturing the qualitative features and even most of the quantitative features of the initial response functions. In particular, $\chi^{\prime\prime}_{E_xT_x}(\omega)$ has a region of negative spectral weight that is remarkably well reproduced by our approach.

\begin{figure}[b!]
	\begin{minipage}{0.46\linewidth}
		\subfloat[]{\includegraphics[width=0.93\textwidth]{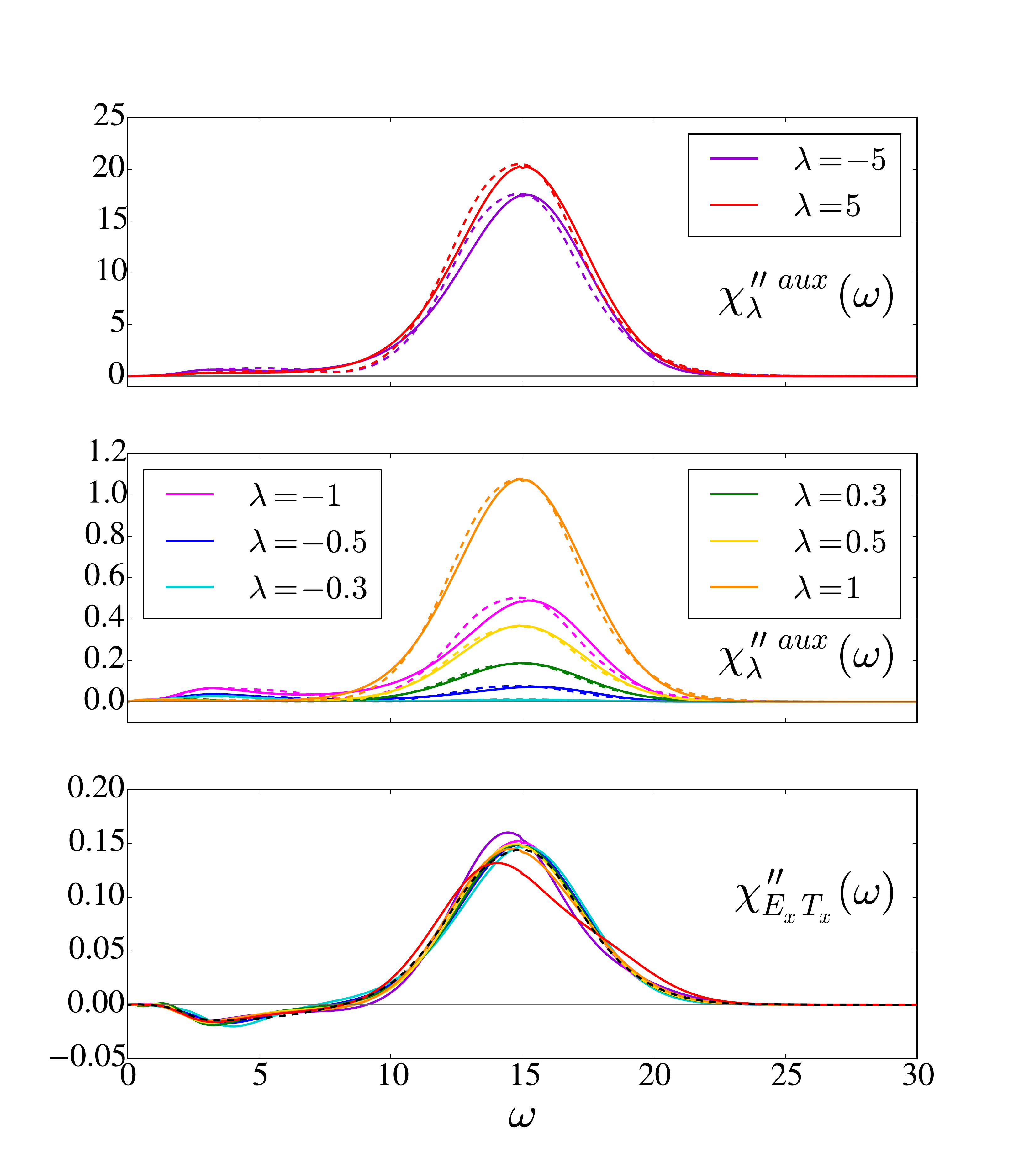} \label{Fig_lambda_Im_Chi_beta10_sup}} 
	\end{minipage}
	\begin{minipage}{0.46\linewidth}
		\subfloat[]{\includegraphics[width=0.92\textwidth]{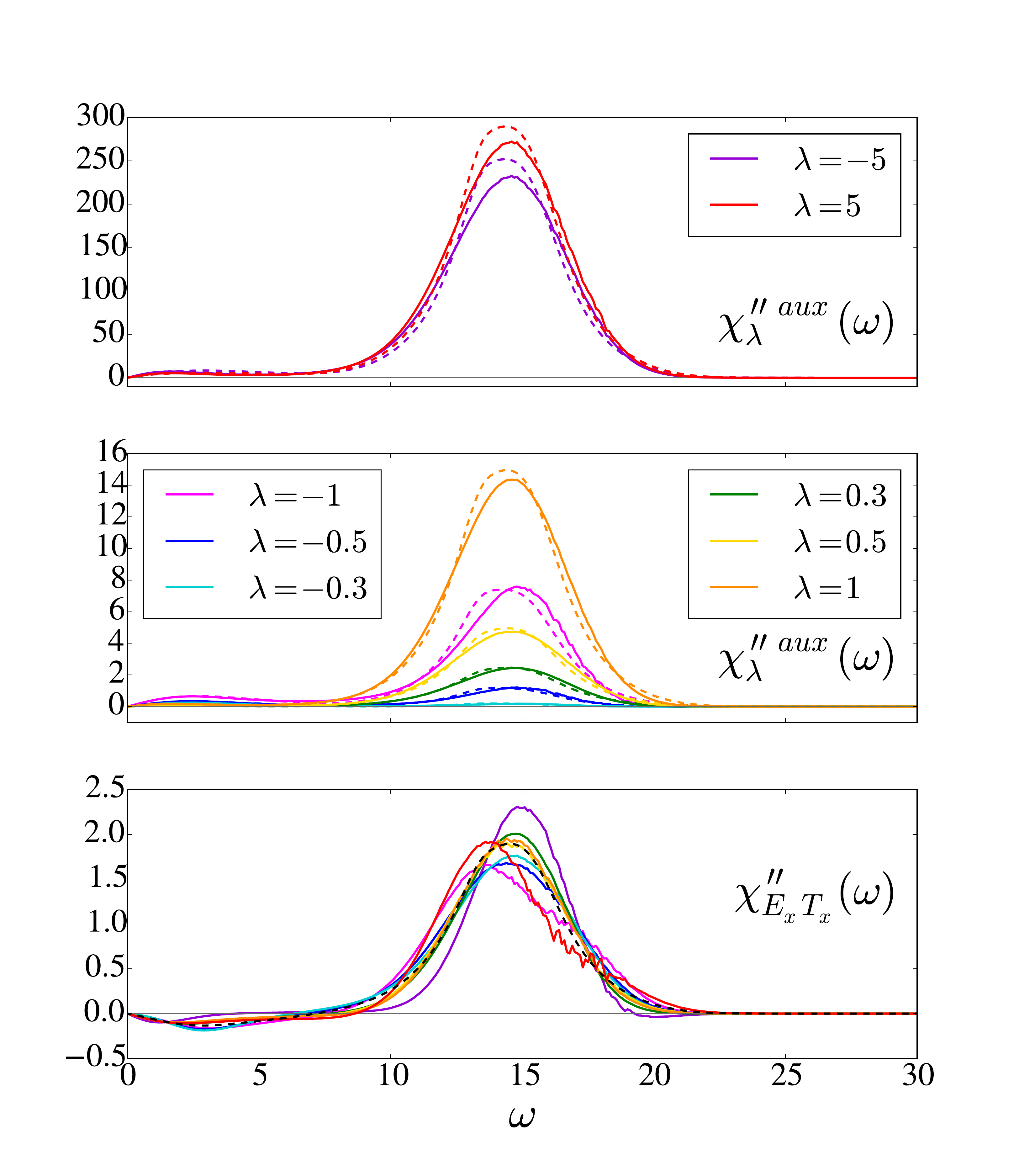} \label{Fig_lambda_Im_Chi_beta1_sup}}
	\end{minipage}
	\centering
	\caption{Comparison of the auxiliary and thermoelectric spectral weights obtained for different values of $\lambda$ from Eq.~\eqref{Eq_Trick_uniform_sup}. The benchmark results before (dashed lines) and the results after the MEACs and the MaxEntAux method (solid lines) are shown. The black dashed lines in the lowest panels represent the expected thermoelectric spectral weight obtained from Eq.~\eqref{Analytic_expression_chi_sup}. (a) Results at $T=1/10$. (b) Results at $T=1$.}
	\label{Fig_lambda_Im_Chi_sup}
\end{figure}

As for Fig.~\ref{Fig_lambda_Im_Chi_beta10_sup} and Fig.~\ref{Fig_lambda_Im_Chi_beta1_sup}, they compare the auxiliary and thermoelectric spectral weights obtained for different values of $\lambda$ from Eq.~\eqref{Eq_Trick_uniform_sup}. They demonstrate the overall robustness of the MaxEntAux method applied for different values of $\lambda$ since most of finite-frequency features remain well preserved. The method becomes noisier at high temperature since MEACs are more difficult in that regime. As explained in the main text, large negative values and small values of $\lambda$ should be avoided. Discrepancies associated with those ranges of $\lambda$ are more apparent in the Onsager coefficients at low frequencies because of the division by $\omega$.

\section{More tests on the Real-frequency and Zero-frequency Onsager Coefficients.} \label{Sec:More_tests} The MaxEntAux method also allows to extract the  frequency-dependent Onsager coefficient usually unobtainable by MEAC, 
\begin{equation}
\mathcal{L}_{E_x T_x}(\omega) = \frac{1}{2\lambda} \left[ \mathcal{L}^{aux}_\lambda(\omega) - \mathcal{L}_{E_x E_x}(\omega) - \lambda^2 \mathcal{L}_{T_x T_x}(\omega) \right]\, ,
\label{Eq_Trick_uniform_L_sup}
\end{equation}
by using the auxiliary functions as in Eq.~\eqref{Eq_Trick_uniform_sup}. Fig.~\ref{Fig_L_beta10_sup} and Fig.~\ref{Fig_L_beta1_sup} compare the benchmark real-frequency Onsager coefficients Eq.~\eqref{Analytic_expression_chi_sup}, at $U=14$, to those obtained with  Eq.~\eqref{Eq_Trick_uniform_L_sup} above and direct MEACs of the Matsubara-frequency data,  with an added relative noise of $10^{-3}$. The analytic expression  Eq.~\eqref{Eq_Onsager_coefficients_sup} that relates those coefficients  to the zero-frequency limit of susceptibilities has also been used.

\begin{figure}[h!]
	\begin{minipage}{0.47\linewidth}
		\subfloat[]{\includegraphics[width=0.92\textwidth]{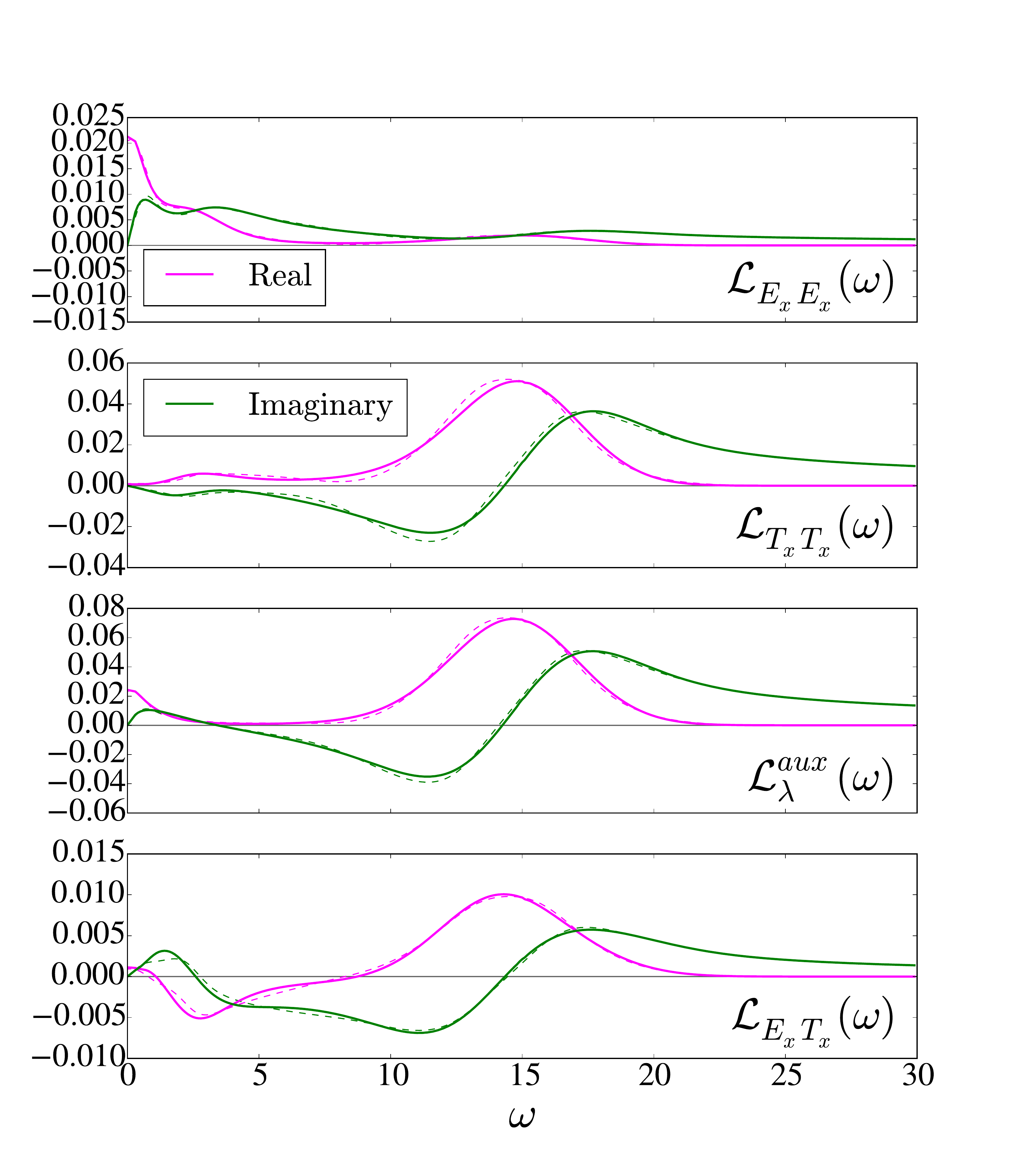} \label{Fig_L_beta10_sup}} 
	\end{minipage}
	\begin{minipage}{0.47\linewidth}
		\subfloat[]{\includegraphics[width=0.9\textwidth]{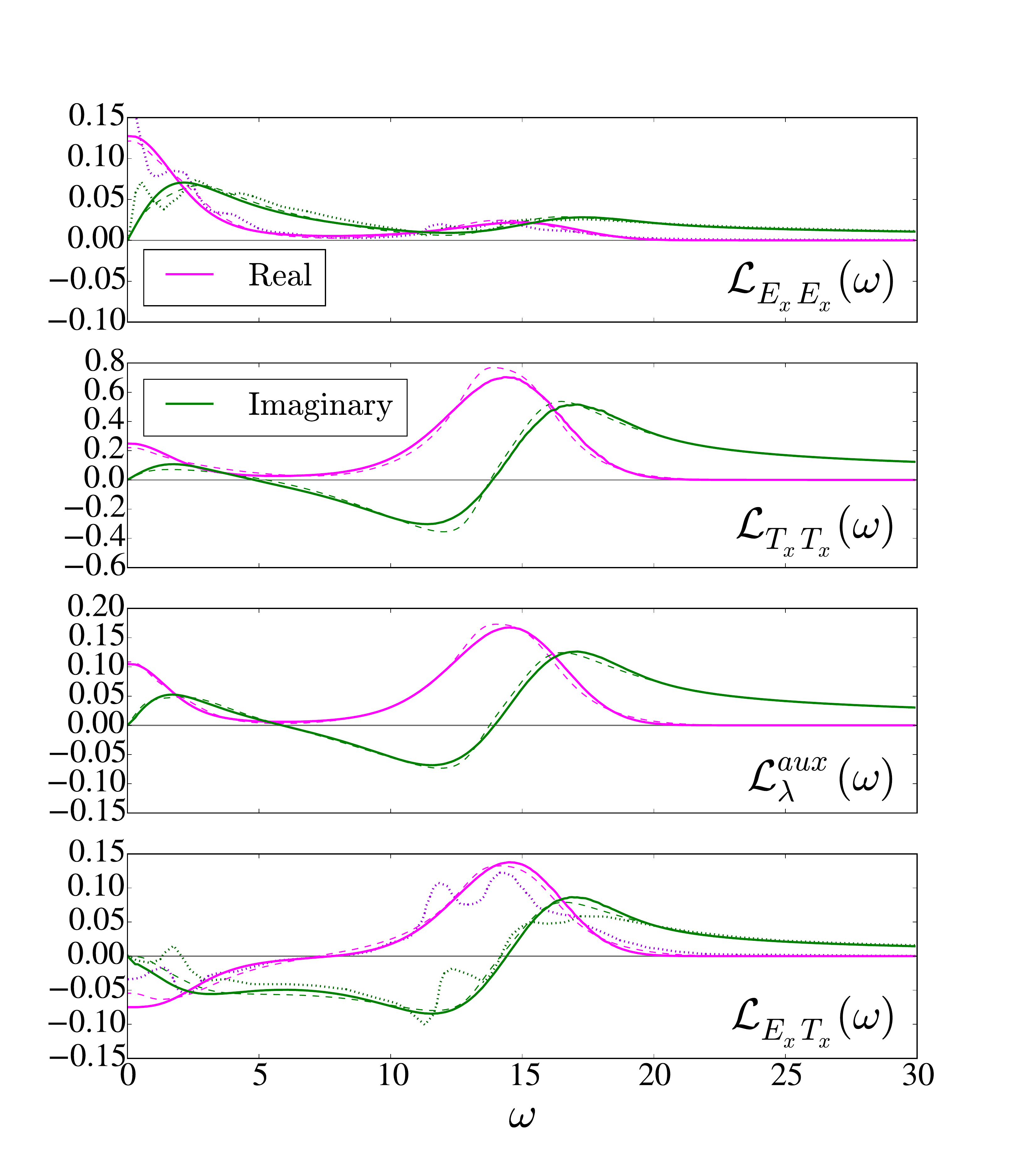} \label{Fig_L_beta1_sup}}
	\end{minipage}
	\centering
	\caption{Comparison between the initial real-frequency Onsager coefficients extracted from the analytic expression Eq.~\eqref{Analytic_expression_chi_sup} using Eq.~\eqref{Eq_Onsager_coefficients_sup} (dashed lines) to those obtained after direct MEACs of the Matsubara-frequency data and the MaxEntAux method Eq.~\eqref{Eq_Trick_uniform_L_sup} (solid lines). (a) Results for $\lambda=1$ at $T=1/10$. (b) Results for $\lambda=0.3$ at $T=1$. The dotted lines show the results of Ref.~[\onlinecite{Xu:2011}] at $T=1$ for comparison.}
	\label{Fig_L_sup}
\end{figure}

In MaxEntAux method Eq.~\eqref{Eq_Trick_uniform_L_sup}, we take  $\lambda=1$ at $T=1/10$ and $\lambda=0.3$ at $T=1$, respectively. Once again, the MaxEntAux method proves very efficient at capturing the qualitative features, and even most of the quantitative features, of the initial frequency-dependent Onsager coefficients. However, the MEACs tend to slightly oscillate around the initial Onsager coefficients obtained from Eq.~\eqref{Analytic_expression_chi_sup} since the almost imperceptible oscillations of the analytically continued spectral weights are amplified by the division by $\omega$ necessary to obtain the Onsager coefficients in Eq.~\eqref{Eq_Onsager_coefficients_sup}. The dotted lines of Fig.~\ref{Fig_L_beta1_sup} shows the exact diagonalization results of Ref.~[\onlinecite{Xu:2011}], which compare well with our results. The latter comparison is provided to show that our results obtained from the combination of CDMFT with continuous-time quantum Monte Carlo (CTQMC) and MEAC are close to those obtained from CDMFT with exact diagonalization (ED). This comparison should not be considered a test of the analytic continuation procedure by itself. As usual, ED finds more numerous and sharper structures than in reality because of its finite bath, whereas CTQMC+MEAC tends to smooth them out.

\begin{figure}[h!]
	\begin{minipage}{0.495\linewidth}
		\subfloat[]{\includegraphics[width=0.84\textwidth]{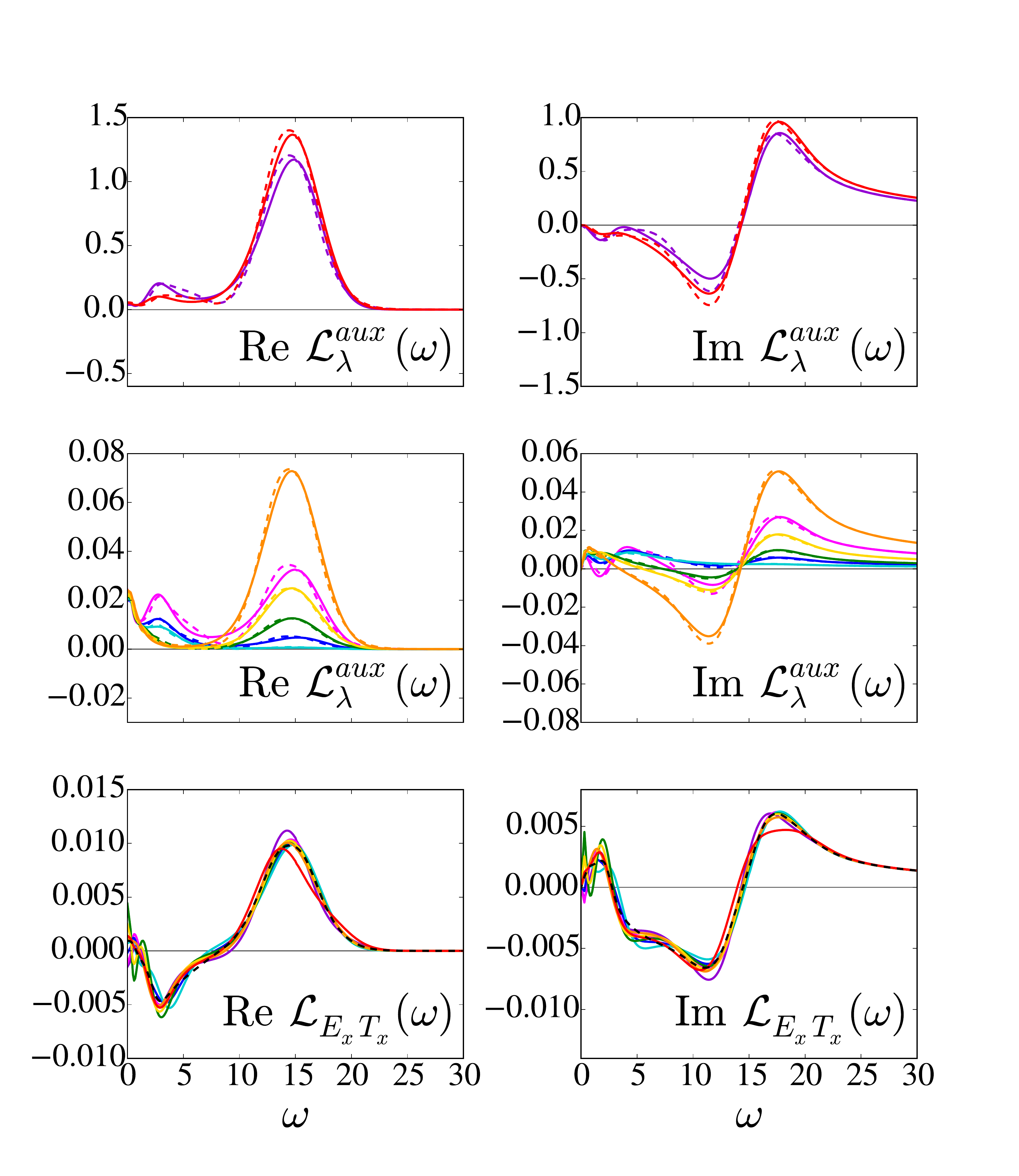} \label{Fig_lambda_L_beta10_sup}} 
	\end{minipage}
	\begin{minipage}{0.495\linewidth}
		\subfloat[]{\includegraphics[width=0.82\textwidth]{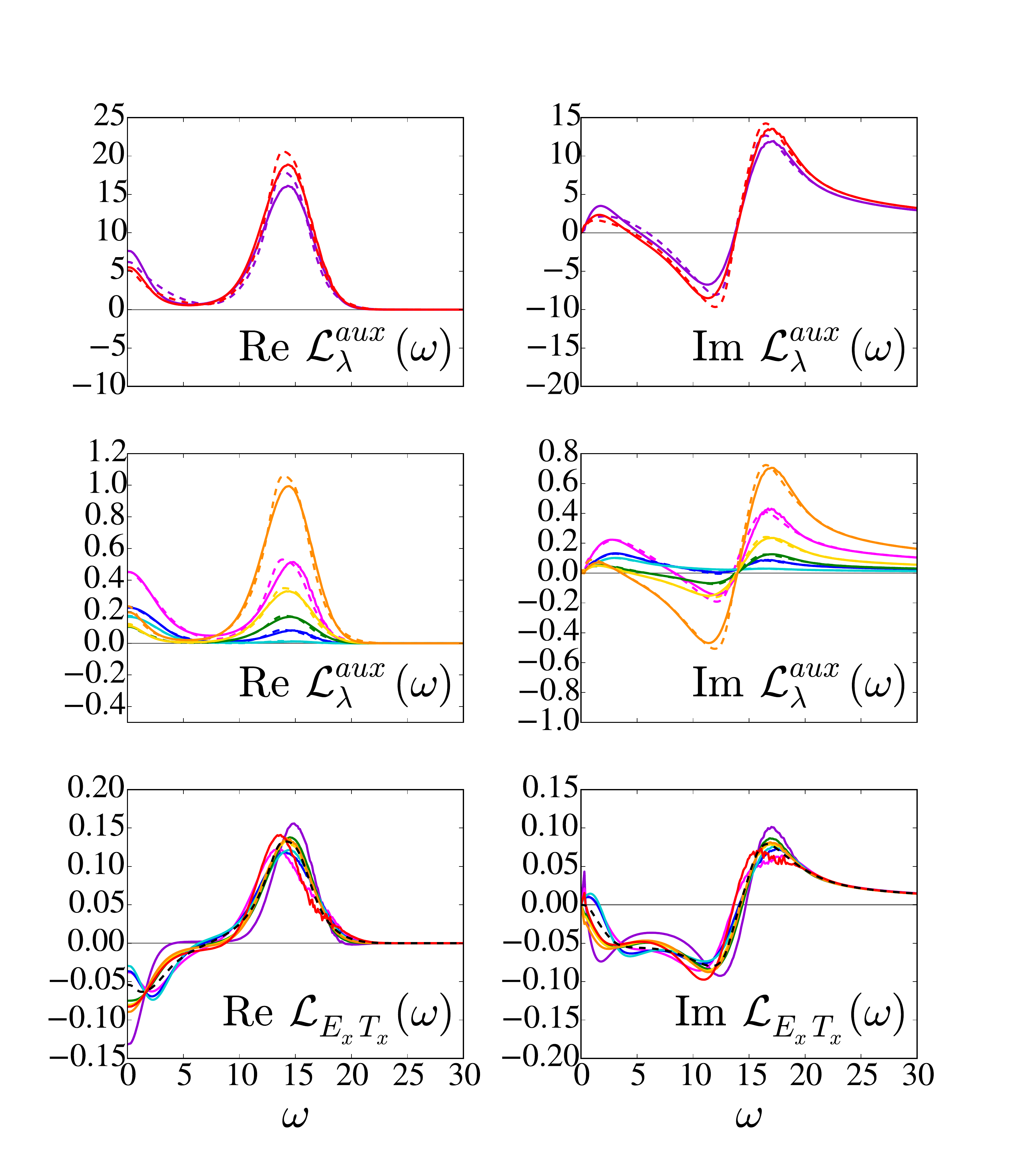} \label{Fig_lambda_L_beta1_sup}}
	\end{minipage}
	\centering
	\caption{Comparison of the auxiliary and thermoelectric frequency-dependent Onsager coefficients obtained for different values of $\lambda$ from Eq.~\eqref{Eq_Trick_uniform_L_sup}. The results before (dashed lines) and after (solid lines) the MEACs and the MaxEntAux method are shown. The color code is the same as the one of Fig.~\ref{Fig_lambda_Im_Chi_sup}. The imaginary part of the Onsager coefficients is obtained from Kramers-Kronig. The black dashed lines in the lowest panels represent the expected thermoelectric frequency-dependent Onsager coefficient obtained from Eq.~\eqref{Analytic_expression_chi_sup} using Eq.~\eqref{Eq_Onsager_coefficients_sup}. (a) Results at $T=1/10$. (b) Results at $T=1$.}
	\label{Fig_lambda_L_sup}
\end{figure}

Fig.~\ref{Fig_lambda_L_beta10_sup} and Fig.~\ref{Fig_lambda_L_beta1_sup} compare the auxiliary and thermoelectric frequency-dependent Onsager coefficients obtained from Eq.~\eqref{Eq_Trick_uniform_sup} for different values of $\lambda$. In the same way as above for the relevant susceptibilities, they demonstrate the overall robustness of the MaxEntAux method applied for different values of $\lambda$. For reasons explained in the main text, we expect that the results are more reliable for $\lambda > 0.5$. The results of Fig.~\ref{Fig_lambda_L_beta1_sup} at $T=1$ are the same as in the main text. They are provided to ease the comparison with $T=1/10$. As expected, analytic continuation is easier at lower temperature.

\begin{figure}[h!]
	\begin{minipage}{0.46\linewidth}
		\subfloat[]{\includegraphics[width=0.93\textwidth]{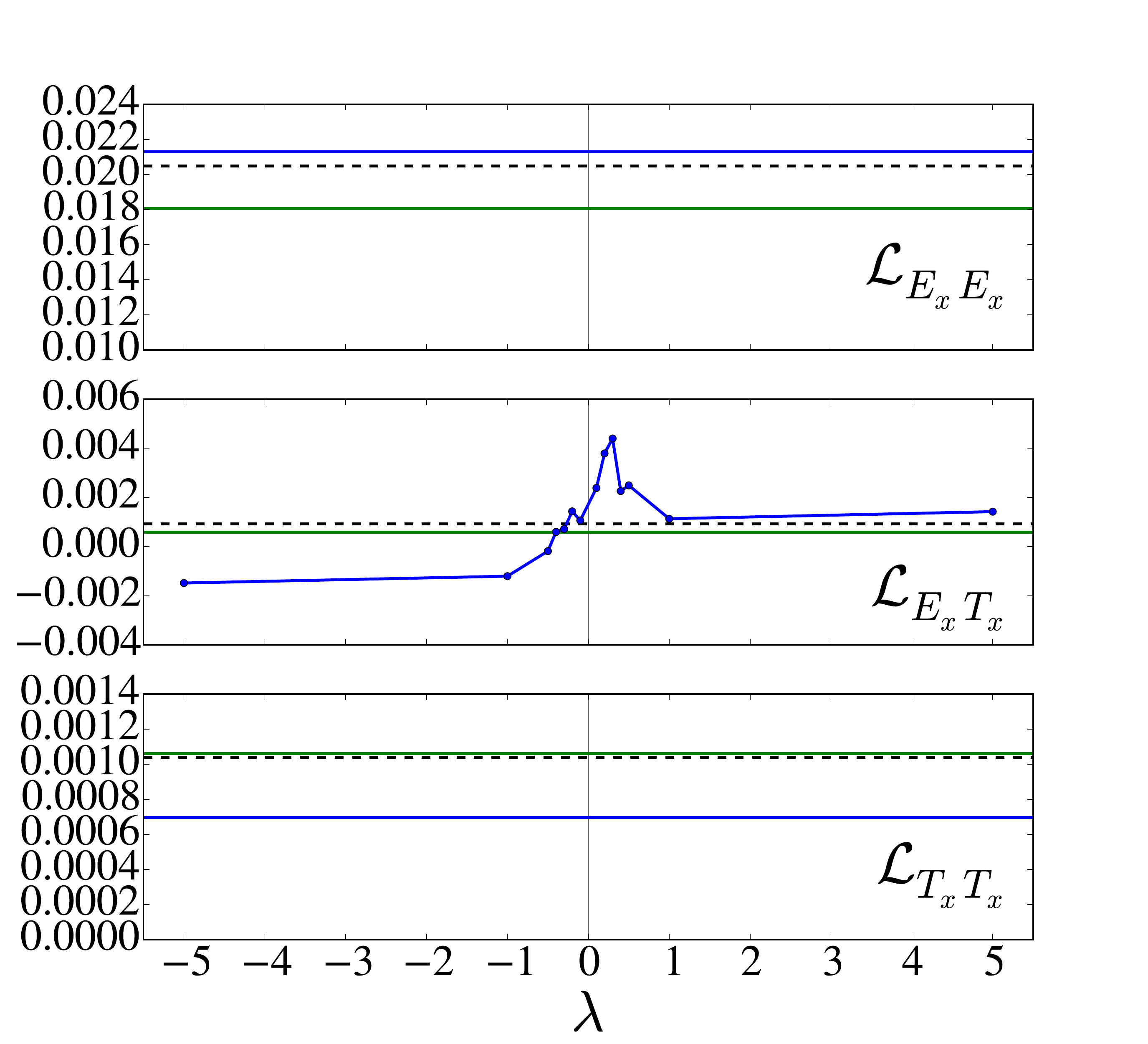} \label{Fig_Onsager_beta10_sup}} 
	\end{minipage}
	\begin{minipage}{0.46\linewidth}
		\subfloat[]{\includegraphics[width=0.92\textwidth]{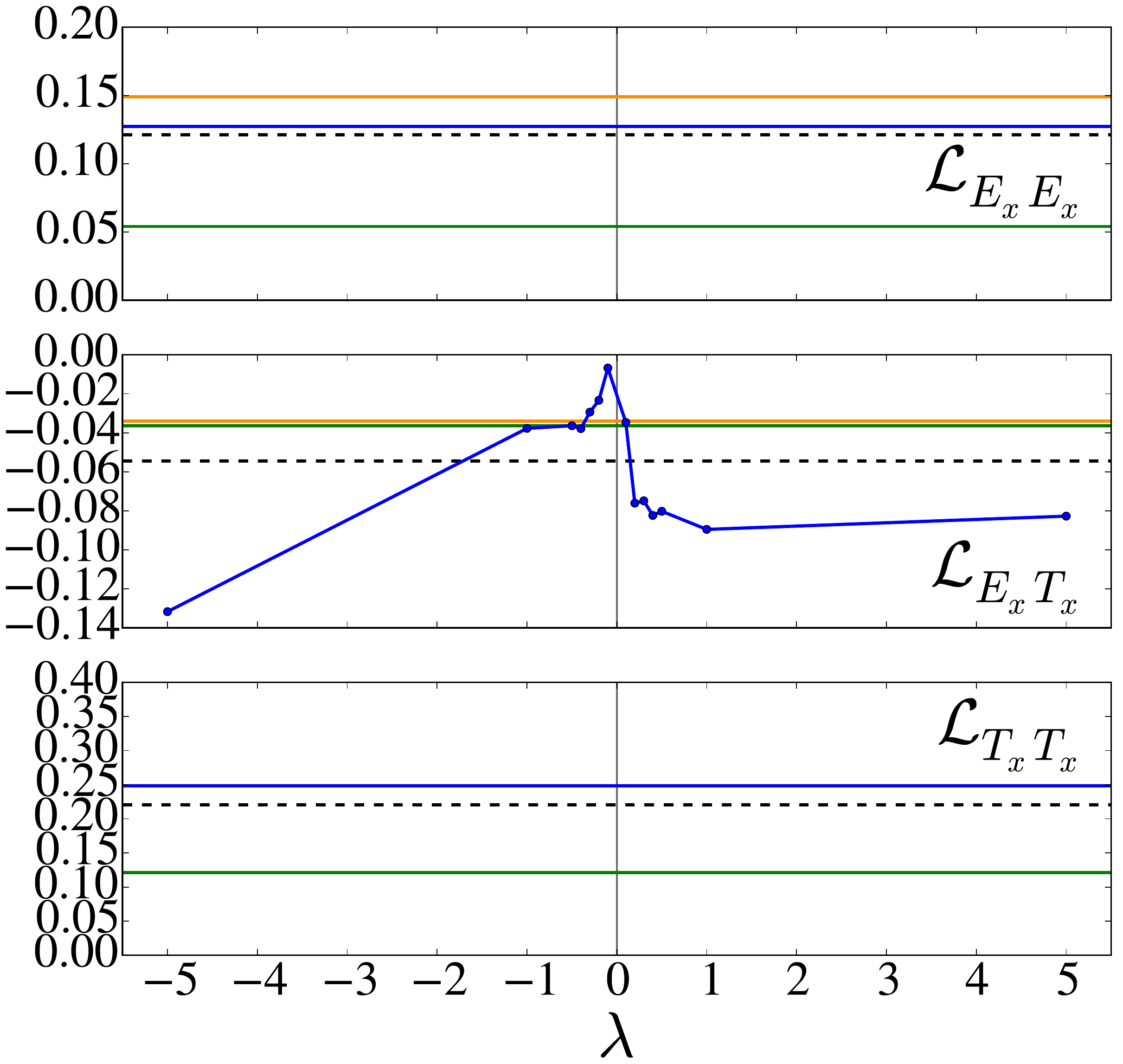} \label{Fig_Onsager_beta1_sup}}
	\end{minipage}
	\centering
	\caption{Zero-frequency values of the frequency-dependent Onsager coefficients shown in Fig.~\ref{Fig_L_sup}. The comparison is made between the results before (black dashed lines) and after (blue solid line in top and bottom panels, and blue points in other panels) the MEACs and the MaxEntAux method for different values of $\lambda$. The blue solid lines in the panels showing $\mathcal{L}_{E_xT_x}$ are guides to the eye for the associated $\lambda$-dependent results (blue points). The green lines show the results of the low-temperature approximation Eq.~\eqref{Eq_L_Randeria_sup}. (a) Results at $T=1/10$. (b) Results at $T=1$. The orange lines show the results of Ref.~[\onlinecite{Xu:2011}] at $T=1$ for consistency of CDMFT, not as a check of analytic continuation.}
	\label{Fig_Onsager_sup}
\end{figure}

Fig.~\ref{Fig_Onsager_beta10_sup} and Fig.~\ref{Fig_Onsager_beta1_sup} compare only the zero-frequency values of the aforementioned frequency-dependent Onsager coefficients, $\mathcal{L}_{\gamma\delta}$, obtained before and after the MEAC, for $T=1/10$ and $T=1$, ($U=14$) respectively. In addition to the comparison of our results with those of Ref.~[\onlinecite{Xu:2011}] for $T=1$ (orange lines), a comparison with the low-temperature approximation, Eq.~\eqref{Eq_L_Randeria_sup}, (green lines) is also shown for both temperatures. (The comparison with the results of Ref.~[\onlinecite{Xu:2011}] that were obtained directly in real-frequency with an ED solver is provided only as a check of the overall consistency of CDMFT, independently of the impurity solver, not as a test of the validity of MaxEntAux.) While the results obtained in Fig.~\ref{Fig_Onsager_beta10_sup} and Fig.~\ref{Fig_Onsager_beta1_sup} with the approximation Eq.~\eqref{Eq_L_Randeria_sup} (green lines) are in better agreement with the benchmark result (black dashed lines) at the lower temperature shown in Fig.~\ref{Fig_Onsager_beta10_sup}, the MaxEntAux method is still competitive. Besides, Eq.~\eqref{Eq_L_Randeria_sup} would not be valid if the system was in the renormalized-classical regime and there is no approximation analog to Eq.~\eqref{Eq_L_Randeria_sup} that can be used for finite-frequency transport quantities to avoid direct analytic continuation. 

The MaxEntAux method seems to be a robust and accurate method for the extraction of the relevant transport coefficients, as long as one uses values of $\lambda$ in a range where the results are $\lambda$-independent and where $\lambda$ is not too close to zero, as discussed in the main text. The Seebeck coefficient shown in Fig.~\ref{Fig_lambda_Im_Chi_beta1_sup} of the main text for $T=1$ and (discussed further below) corresponds to the ratio of Onsager coefficients that is usually measured. It is more accurate than the value of  $\mathcal{L}_{E_xT_x}$ shown here.


\newpage

\begin{figure}[h!]
	\begin{minipage}{0.49\linewidth}
		\subfloat[]{\includegraphics[width=0.84\textwidth]{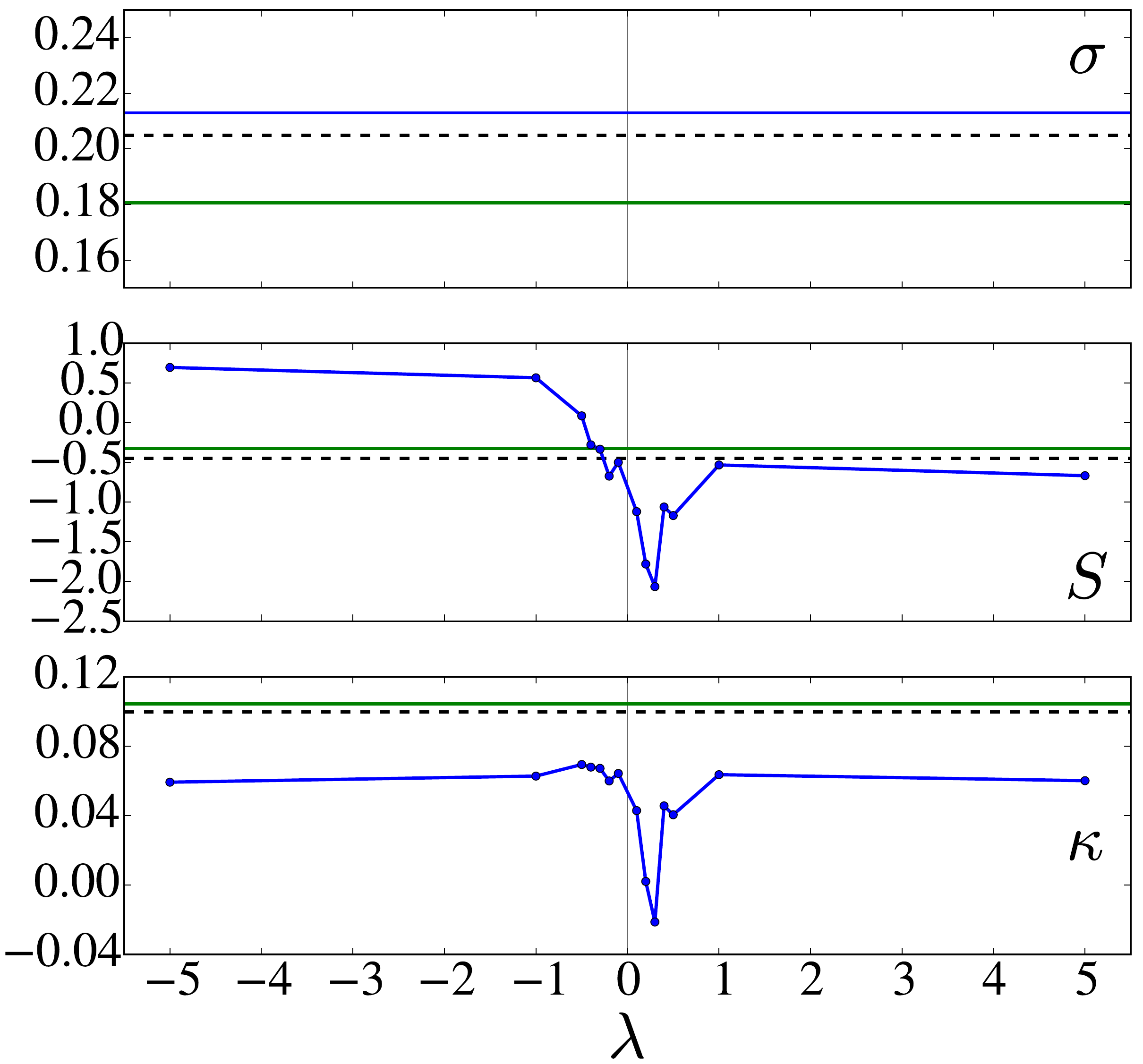} \label{Fig_sigma_S_beta10_sup}} 
	\end{minipage}
	\begin{minipage}{0.49\linewidth}
		\subfloat[]{\includegraphics[width=0.82\textwidth]{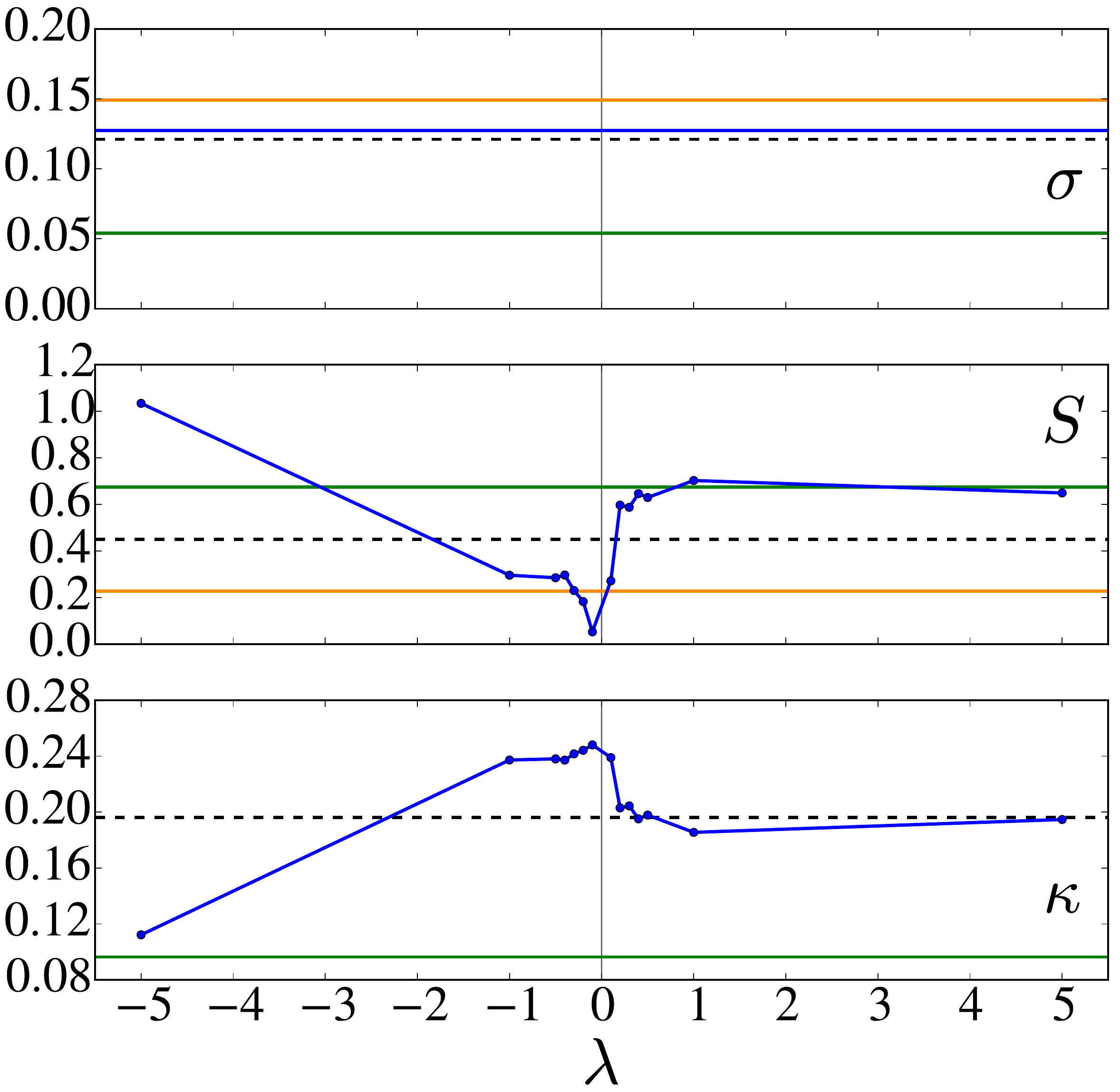} \label{Fig_sigma_S_beta1_sup}}
	\end{minipage}
	\centering
	\caption{Electrical conductivity, Seebeck coefficient, and thermal conductivity extracted from Eq.~\eqref{Eq_sigma_sup} and Eq.~\eqref{Eq_S_sup}, and Eq.~\eqref{Eq_kappa_sup}, respectively. The comparison is made between the results before (black dashed lines) and after (blue solid line in top panels and blue points in other panels) the MEACs and the MaxEntAux method for different values of $\lambda$. The blue solid lines in the panels showing $S$ and $\kappa$ are guides to the eye for the associated $\lambda$-dependent results (blue points). The green lines show the results of the low-temperature approximation Eq.~\eqref{Eq_L_Randeria_sup}. (a) Results at $T=1/10$. (b) Results at $T=1$. The orange lines show the results of Ref.~[\onlinecite{Xu:2011}] at $T=1$ for comparison.}
	\label{Fig_sigma_S_sup}
\end{figure}

\section{Benchmarks for lower temperature transport coefficients.}\label{Sec:Transport-Coeff} Fig.~\ref{Fig_sigma_S_beta10_sup} and Fig.~\ref{Fig_sigma_S_beta1_sup} show the electrical conductivity $\sigma$ Eq.~\eqref{Eq_sigma_sup}, the Seebeck coefficient $S$ Eq.~\eqref{Eq_S_sup} and the thermal conductivity $\kappa$ Eq.~\eqref{Eq_kappa_sup} at $T=1/10$ and at $T=1$, respectively, for $U=14$. The $T=1$ results appear in the second figure of the main text. They are provided here to ease comparison with lower temperature results. 

With the usual caveats on $\lambda$, the MaxEntAux method proves competitive once more, except for the thermal conductivity at $T=1/10$ where the low-temperature approximation Eq.~\eqref{Eq_L_Randeria_sup} is better. We explain this discrepancy by the sensitivity of Eq.~\eqref{Eq_kappa_sup} for $\kappa$ to errors on the value of the Onsager coefficient $\mathcal{L}_{T_xT_x}$, which is difficult to analytically continue. Indeed, we saw in Fig.~\ref{Fig_Onsager_beta10_sup} that it was off. Limitations of the low temperature approximation Eq.~\eqref{Eq_L_Randeria_sup} are discussed at the end of the previous section.

\section{Convergence of Matsubara frequency sums: New expressions for the uniform susceptibilities $\chi_{E_xT_x}(i\omega_n)$ and $\chi_{T_xT_x}(i\omega_n)$ without vertex corrections.}\label{Sec:ConvergenceMatsubara} The calculation of Matsubara-frequency susceptibilities, that are input to the MaxEntAux method, is difficult by itself. In particular, the expressions given in Ref.~[\onlinecite{Paul:2003}] for the thermoelectric and thermal conductivity transport coefficients lack the terms that make the summations over Matsubara frequencies ($ip_m$ below) convergent. The Master's thesis of A.-M. Gagnon Ref.~[\onlinecite{GagnonMSc:2016}] gives us new convergent expressions for the bubble contribution to $\chi_{E_xT_x} (i\omega_n)$ and $\chi_{T_xT_x} (i\omega_n)$. Convergence of the terms containing vertex corrections should be easier to obtain. They are not discussed here. Restoring the electrical charge $e$, her results are summarized here:
\begin{eqnarray}
\chi_{E_xE_x} (i\omega_n) & = & 
-\frac{e^2}{\beta}\sum_{\vec{k},\sigma} \left( \frac{\partial \varepsilon_{\vec{k}}}{\partial k_x} \right)^2 \sum_{ip_m} \mathcal{G}_\sigma(\vec{k},ip_m) \mathcal{G}_\sigma(\vec{k},ip_m + i\omega_n)\, , \\
\chi_{E_xT_x} (i\omega_n) & = & 
-\frac{e}{\beta}\sum_{\vec{k},\sigma} \left( \frac{\partial \varepsilon_{\vec{k}}}{\partial k_x} \right)^2 \sum_{ip_m} \left[ \left( ip_m + \frac{i\omega_n}{2} \right) \mathcal{G}_\sigma(\vec{k},ip_m) \mathcal{G}_\sigma(\vec{k},ip_m + i\omega_n) - \mathcal{G}_\sigma(\vec{k},ip_m) \right] , \\
\chi_{T_xT_x} (i\omega_n) & = & 
-\frac{1}{4\beta}\sum_{\vec{k},\sigma} \left( \frac{\partial \varepsilon_{\vec{k}}}{\partial k_x} \right)^2 \sum_{ip_m} \left[ -(\langle\omega\rangle_\sigma(\vec{k}) - ip_m) \mathcal{G}_\sigma(\vec{k},ip_m + i\omega_n) + 2 - (\langle\omega\rangle_\sigma(\vec{k}) - ip_m - i\omega_n)\mathcal{G}_\sigma(\vec{k},ip_m) \right. \nonumber \\
 & & \left. - 2(2ip_m + i\omega_n) \left( \mathcal{G}_\sigma(\vec{k},ip_m + i\omega_n) + \mathcal{G}_\sigma(\vec{k},ip_m) \right) + (2ip_m + i\omega_n)^2 \mathcal{G}_\sigma(\vec{k},ip_m) \mathcal{G}_\sigma(\vec{k},ip_m + i\omega_n) \right] ,
\end{eqnarray}
where, in the case of the Hubbard model, 
\begin{equation}
\langle\omega\rangle_\sigma(\vec{k}) = \int \! \omega\, \mathcal{A}_\sigma(\vec{k},\omega)\, \mathrm{d}\omega = \varepsilon_{\vec{k}} - \mu +Un_{\sigma} \, .
\end{equation}
$n_{\sigma}$ represents the filling and $\mathcal{A}_\sigma(\vec{k},\omega)$ the spectral weight of electrons with spin $\sigma$. 

Although the above sums are convergent, for large $i\omega_n$ one needs to be careful. Convergence can be accelerated greatly by using some of the algorithms that appear in section C of appendix C of Ref.~[\onlinecite{Bergeron:2011}].

\end{widetext}


\begin{thebibliography}{43}%
\makeatletter
\providecommand \@ifxundefined [1]{%
 \@ifx{#1\undefined}
}%
\providecommand \@ifnum [1]{%
 \ifnum #1\expandafter \@firstoftwo
 \else \expandafter \@secondoftwo
 \fi
}%
\providecommand \@ifx [1]{%
 \ifx #1\expandafter \@firstoftwo
 \else \expandafter \@secondoftwo
 \fi
}%
\providecommand \natexlab [1]{#1}%
\providecommand \enquote  [1]{``#1''}%
\providecommand \bibnamefont  [1]{#1}%
\providecommand \bibfnamefont [1]{#1}%
\providecommand \citenamefont [1]{#1}%
\providecommand \href@noop [0]{\@secondoftwo}%
\providecommand \href [0]{\begingroup \@sanitize@url \@href}%
\providecommand \@href[1]{\@@startlink{#1}\@@href}%
\providecommand \@@href[1]{\endgroup#1\@@endlink}%
\providecommand \@sanitize@url [0]{\catcode `\\12\catcode `\$12\catcode
  `\&12\catcode `\#12\catcode `\^12\catcode `\_12\catcode `\%12\relax}%
\providecommand \@@startlink[1]{}%
\providecommand \@@endlink[0]{}%
\providecommand \url  [0]{\begingroup\@sanitize@url \@url }%
\providecommand \@url [1]{\endgroup\@href {#1}{\urlprefix }}%
\providecommand \urlprefix  [0]{URL }%
\providecommand \Eprint [0]{\href }%
\providecommand \doibase [0]{http://dx.doi.org/}%
\providecommand \selectlanguage [0]{\@gobble}%
\providecommand \bibinfo  [0]{\@secondoftwo}%
\providecommand \bibfield  [0]{\@secondoftwo}%
\providecommand \translation [1]{[#1]}%
\providecommand \BibitemOpen [0]{}%
\providecommand \bibitemStop [0]{}%
\providecommand \bibitemNoStop [0]{.\EOS\space}%
\providecommand \EOS [0]{\spacefactor3000\relax}%
\providecommand \BibitemShut  [1]{\csname bibitem#1\endcsname}%
\let\auto@bib@innerbib\@empty
\bibitem [{\citenamefont {Terasaki}\ \emph {et~al.}(1997)\citenamefont
  {Terasaki}, \citenamefont {Sasago},\ and\ \citenamefont
  {Uchinokura}}]{Terasaki:1997}%
  \BibitemOpen
  \bibfield  {author} {\bibinfo {author} {\bibfnamefont {I.}~\bibnamefont
  {Terasaki}}, \bibinfo {author} {\bibfnamefont {Y.}~\bibnamefont {Sasago}}, \
  and\ \bibinfo {author} {\bibfnamefont {K.}~\bibnamefont {Uchinokura}},\
  }\bibfield  {title} {\enquote {\bibinfo {title} {Large thermoelectric power
  in ${\mathrm{naco}}_{2}{\mathrm{o}}_{4}$ single crystals},}\ }\href {\doibase
  10.1103/PhysRevB.56.R12685} {\bibfield  {journal} {\bibinfo  {journal} {Phys.
  Rev. B}\ }\textbf {\bibinfo {volume} {56}},\ \bibinfo {pages}
  {R12685--R12687} (\bibinfo {year} {1997})}\BibitemShut {NoStop}%
\bibitem [{\citenamefont {Bentien}\ \emph {et~al.}(2007)\citenamefont
  {Bentien}, \citenamefont {Johnsen}, \citenamefont {Madsen}, \citenamefont
  {Iversen},\ and\ \citenamefont {Steglich}}]{Bentien:2007}%
  \BibitemOpen
  \bibfield  {author} {\bibinfo {author} {\bibfnamefont {A.}~\bibnamefont
  {Bentien}}, \bibinfo {author} {\bibfnamefont {S.}~\bibnamefont {Johnsen}},
  \bibinfo {author} {\bibfnamefont {G.~K.~H.}\ \bibnamefont {Madsen}}, \bibinfo
  {author} {\bibfnamefont {B.~B.}\ \bibnamefont {Iversen}}, \ and\ \bibinfo
  {author} {\bibfnamefont {F.}~\bibnamefont {Steglich}},\ }\bibfield  {title}
  {\enquote {\bibinfo {title} {Colossal seebeck coefficient in strongly
  correlated semiconductor fesb 2},}\ }\href
  {http://stacks.iop.org/0295-5075/80/i=1/a=17008} {\bibfield  {journal}
  {\bibinfo  {journal} {EPL (Europhysics Letters)}\ }\textbf {\bibinfo {volume}
  {80}},\ \bibinfo {pages} {17008} (\bibinfo {year} {2007})}\BibitemShut
  {NoStop}%
\bibitem [{\citenamefont {Tomczak}\ \emph {et~al.}(2010)\citenamefont
  {Tomczak}, \citenamefont {Haule}, \citenamefont {Miyake}, \citenamefont
  {Georges},\ and\ \citenamefont {Kotliar}}]{Tomczak:2010}%
  \BibitemOpen
  \bibfield  {author} {\bibinfo {author} {\bibfnamefont {Jan~M.}\ \bibnamefont
  {Tomczak}}, \bibinfo {author} {\bibfnamefont {K.}~\bibnamefont {Haule}},
  \bibinfo {author} {\bibfnamefont {T.}~\bibnamefont {Miyake}}, \bibinfo
  {author} {\bibfnamefont {A.}~\bibnamefont {Georges}}, \ and\ \bibinfo
  {author} {\bibfnamefont {G.}~\bibnamefont {Kotliar}},\ }\bibfield  {title}
  {\enquote {\bibinfo {title} {Thermopower of correlated semiconductors:
  Application to ${\text{feas}}_{2}$ and ${\text{fesb}}_{2}$},}\ }\href
  {\doibase 10.1103/PhysRevB.82.085104} {\bibfield  {journal} {\bibinfo
  {journal} {Phys. Rev. B}\ }\textbf {\bibinfo {volume} {82}},\ \bibinfo
  {pages} {085104} (\bibinfo {year} {2010})}\BibitemShut {NoStop}%
\bibitem [{\citenamefont {Sun}\ \emph {et~al.}(2011)\citenamefont {Sun},
  \citenamefont {Søndergaard}, \citenamefont {Iversen},\ and\ \citenamefont
  {Steglich}}]{Sun:2011}%
  \BibitemOpen
  \bibfield  {author} {\bibinfo {author} {\bibfnamefont {P.}~\bibnamefont
  {Sun}}, \bibinfo {author} {\bibfnamefont {M.}~\bibnamefont {Søndergaard}},
  \bibinfo {author} {\bibfnamefont {B.B.}\ \bibnamefont {Iversen}}, \ and\
  \bibinfo {author} {\bibfnamefont {F.}~\bibnamefont {Steglich}},\ }\bibfield
  {title} {\enquote {\bibinfo {title} {Strong electron correlations in
  fesb2},}\ }\href {\doibase 10.1002/andp.201100033} {\bibfield  {journal}
  {\bibinfo  {journal} {Annalen der Physik}\ }\textbf {\bibinfo {volume}
  {523}},\ \bibinfo {pages} {612--620} (\bibinfo {year} {2011})}\BibitemShut
  {NoStop}%
\bibitem [{\citenamefont {Arsenault}\ \emph {et~al.}(2013)\citenamefont
  {Arsenault}, \citenamefont {Shastry}, \citenamefont {S{\'e}mon},\ and\
  \citenamefont {Tremblay}}]{Arsenault:2013}%
  \BibitemOpen
  \bibfield  {author} {\bibinfo {author} {\bibfnamefont {Louis-Fran{\c c}ois}\
  \bibnamefont {Arsenault}}, \bibinfo {author} {\bibfnamefont {B.~Sriram}\
  \bibnamefont {Shastry}}, \bibinfo {author} {\bibfnamefont {Patrick}\
  \bibnamefont {S{\'e}mon}}, \ and\ \bibinfo {author} {\bibfnamefont
  {A.-M.~S.}\ \bibnamefont {Tremblay}},\ }\bibfield  {title} {\enquote
  {\bibinfo {title} {Entropy, frustration, and large thermopower of doped mott
  insulators on the fcc lattice},}\ }\href {\doibase
  10.1103/PhysRevB.87.035126} {\bibfield  {journal} {\bibinfo  {journal} {Phys.
  Rev. B}\ }\textbf {\bibinfo {volume} {87}},\ \bibinfo {pages} {035126}
  (\bibinfo {year} {2013})}\BibitemShut {NoStop}%
\bibitem [{\citenamefont {Ozaeta}\ \emph {et~al.}(2014)\citenamefont {Ozaeta},
  \citenamefont {Virtanen}, \citenamefont {Bergeret},\ and\ \citenamefont
  {Heikkil\"a}}]{Ozaeta:2014}%
  \BibitemOpen
  \bibfield  {author} {\bibinfo {author} {\bibfnamefont {A.}~\bibnamefont
  {Ozaeta}}, \bibinfo {author} {\bibfnamefont {P.}~\bibnamefont {Virtanen}},
  \bibinfo {author} {\bibfnamefont {F.~S.}\ \bibnamefont {Bergeret}}, \ and\
  \bibinfo {author} {\bibfnamefont {T.~T.}\ \bibnamefont {Heikkil\"a}},\
  }\bibfield  {title} {\enquote {\bibinfo {title} {Predicted very large
  thermoelectric effect in ferromagnet-superconductor junctions in the presence
  of a spin-splitting magnetic field},}\ }\href {\doibase
  10.1103/PhysRevLett.112.057001} {\bibfield  {journal} {\bibinfo  {journal}
  {Phys. Rev. Lett.}\ }\textbf {\bibinfo {volume} {112}},\ \bibinfo {pages}
  {057001} (\bibinfo {year} {2014})}\BibitemShut {NoStop}%
\bibitem [{\citenamefont {Kolenda}\ \emph {et~al.}(2016)\citenamefont
  {Kolenda}, \citenamefont {Wolf},\ and\ \citenamefont
  {Beckmann}}]{Kolenda:2016}%
  \BibitemOpen
  \bibfield  {author} {\bibinfo {author} {\bibfnamefont {S.}~\bibnamefont
  {Kolenda}}, \bibinfo {author} {\bibfnamefont {M.~J.}\ \bibnamefont {Wolf}}, \
  and\ \bibinfo {author} {\bibfnamefont {D.}~\bibnamefont {Beckmann}},\
  }\bibfield  {title} {\enquote {\bibinfo {title} {Observation of
  thermoelectric currents in high-field superconductor-ferromagnet tunnel
  junctions},}\ }\href {\doibase 10.1103/PhysRevLett.116.097001} {\bibfield
  {journal} {\bibinfo  {journal} {Phys. Rev. Lett.}\ }\textbf {\bibinfo
  {volume} {116}},\ \bibinfo {pages} {097001} (\bibinfo {year}
  {2016})}\BibitemShut {NoStop}%
\bibitem [{\citenamefont {Jarrell}\ and\ \citenamefont
  {Gubernatis}(1996)}]{Jarrell:1996}%
  \BibitemOpen
  \bibfield  {author} {\bibinfo {author} {\bibfnamefont {Mark}\ \bibnamefont
  {Jarrell}}\ and\ \bibinfo {author} {\bibfnamefont {J.~E.}\ \bibnamefont
  {Gubernatis}},\ }\bibfield  {title} {\enquote {\bibinfo {title} {Bayesian
  inference and the analytic continuation of imaginary-time quantum {Monte
  Carlo} data},}\ }\href {\doibase 10.1016/0370-1573(95)00074-7} {\bibfield
  {journal} {\bibinfo  {journal} {Physics Reports}\ }\textbf {\bibinfo {volume}
  {269}},\ \bibinfo {pages} {133--195} (\bibinfo {year} {1996})}\BibitemShut
  {NoStop}%
\bibitem [{\citenamefont {Sandvik}(1998)}]{Sandvik:1998}%
  \BibitemOpen
  \bibfield  {author} {\bibinfo {author} {\bibfnamefont {Anders~W.}\
  \bibnamefont {Sandvik}},\ }\bibfield  {title} {\enquote {\bibinfo {title}
  {Stochastic method for analytic continuation of quantum monte carlo data},}\
  }\href {\doibase 10.1103/PhysRevB.57.10287} {\bibfield  {journal} {\bibinfo
  {journal} {Phys. Rev. B}\ }\textbf {\bibinfo {volume} {57}},\ \bibinfo
  {pages} {10287--10290} (\bibinfo {year} {1998})}\BibitemShut {NoStop}%
\bibitem [{\citenamefont {Beni}(1974)}]{Beni:1974}%
  \BibitemOpen
  \bibfield  {author} {\bibinfo {author} {\bibfnamefont {Gerardo}\ \bibnamefont
  {Beni}},\ }\bibfield  {title} {\enquote {\bibinfo {title} {Thermoelectric
  power of the narrow-band hubbard chain at arbitrary electron density: Atomic
  limit},}\ }\href {\doibase 10.1103/PhysRevB.10.2186} {\bibfield  {journal}
  {\bibinfo  {journal} {Phys. Rev. B}\ }\textbf {\bibinfo {volume} {10}},\
  \bibinfo {pages} {2186--2189} (\bibinfo {year} {1974})}\BibitemShut {NoStop}%
\bibitem [{\citenamefont {Chaikin}\ and\ \citenamefont
  {Beni}(1976)}]{Chaikin:1976}%
  \BibitemOpen
  \bibfield  {author} {\bibinfo {author} {\bibfnamefont {P.~M.}\ \bibnamefont
  {Chaikin}}\ and\ \bibinfo {author} {\bibfnamefont {G.}~\bibnamefont {Beni}},\
  }\bibfield  {title} {\enquote {\bibinfo {title} {Thermopower in the
  correlated hopping regime},}\ }\href {\doibase 10.1103/PhysRevB.13.647}
  {\bibfield  {journal} {\bibinfo  {journal} {Phys. Rev. B}\ }\textbf {\bibinfo
  {volume} {13}},\ \bibinfo {pages} {647--651} (\bibinfo {year}
  {1976})}\BibitemShut {NoStop}%
\bibitem [{\citenamefont {P\'alsson}\ and\ \citenamefont
  {Kotliar}(1998)}]{Palsson:1998}%
  \BibitemOpen
  \bibfield  {author} {\bibinfo {author} {\bibfnamefont {Gunnar}\ \bibnamefont
  {P\'alsson}}\ and\ \bibinfo {author} {\bibfnamefont {Gabriel}\ \bibnamefont
  {Kotliar}},\ }\bibfield  {title} {\enquote {\bibinfo {title} {Thermoelectric
  response near the density driven mott transition},}\ }\href {\doibase
  10.1103/PhysRevLett.80.4775} {\bibfield  {journal} {\bibinfo  {journal}
  {Phys. Rev. Lett.}\ }\textbf {\bibinfo {volume} {80}},\ \bibinfo {pages}
  {4775--4778} (\bibinfo {year} {1998})}\BibitemShut {NoStop}%
\bibitem [{\citenamefont {Koshibae}\ \emph {et~al.}(2000)\citenamefont
  {Koshibae}, \citenamefont {Tsutsui},\ and\ \citenamefont
  {Maekawa}}]{Koshibae:2000}%
  \BibitemOpen
  \bibfield  {author} {\bibinfo {author} {\bibfnamefont {W.}~\bibnamefont
  {Koshibae}}, \bibinfo {author} {\bibfnamefont {K.}~\bibnamefont {Tsutsui}}, \
  and\ \bibinfo {author} {\bibfnamefont {S.}~\bibnamefont {Maekawa}},\
  }\bibfield  {title} {\enquote {\bibinfo {title} {Thermopower in cobalt
  oxides},}\ }\href {\doibase 10.1103/PhysRevB.62.6869} {\bibfield  {journal}
  {\bibinfo  {journal} {Phys. Rev. B}\ }\textbf {\bibinfo {volume} {62}},\
  \bibinfo {pages} {6869--6872} (\bibinfo {year} {2000})}\BibitemShut {NoStop}%
\bibitem [{\citenamefont {Oudovenko}\ and\ \citenamefont
  {Kotliar}(2002)}]{Oudovenko:2002}%
  \BibitemOpen
  \bibfield  {author} {\bibinfo {author} {\bibfnamefont {V.~S.}\ \bibnamefont
  {Oudovenko}}\ and\ \bibinfo {author} {\bibfnamefont {G.}~\bibnamefont
  {Kotliar}},\ }\bibfield  {title} {\enquote {\bibinfo {title} {Thermoelectric
  properties of the degenerate hubbard model},}\ }\href {\doibase
  10.1103/PhysRevB.65.075102} {\bibfield  {journal} {\bibinfo  {journal} {Phys.
  Rev. B}\ }\textbf {\bibinfo {volume} {65}},\ \bibinfo {pages} {075102}
  (\bibinfo {year} {2002})}\BibitemShut {NoStop}%
\bibitem [{\citenamefont {Kontani}(2003)}]{Kontani:2003}%
  \BibitemOpen
  \bibfield  {author} {\bibinfo {author} {\bibfnamefont {Hiroshi}\ \bibnamefont
  {Kontani}},\ }\bibfield  {title} {\enquote {\bibinfo {title} {General formula
  for the thermoelectric transport phenomena based on fermi liquid theory:
  Thermoelectric power, nernst coefficient, and thermal conductivity},}\ }\href
  {\doibase 10.1103/PhysRevB.67.014408} {\bibfield  {journal} {\bibinfo
  {journal} {Phys. Rev. B}\ }\textbf {\bibinfo {volume} {67}},\ \bibinfo
  {pages} {014408} (\bibinfo {year} {2003})}\BibitemShut {NoStop}%
\bibitem [{\citenamefont {Shastry}(2006)}]{Shastry:2006}%
  \BibitemOpen
  \bibfield  {author} {\bibinfo {author} {\bibfnamefont {B.~Sriram}\
  \bibnamefont {Shastry}},\ }\bibfield  {title} {\enquote {\bibinfo {title}
  {Sum rule for thermal conductivity and dynamical thermal transport
  coefficients in condensed matter},}\ }\href {\doibase
  10.1103/PhysRevB.73.085117} {\bibfield  {journal} {\bibinfo  {journal} {Phys.
  Rev. B}\ }\textbf {\bibinfo {volume} {73}},\ \bibinfo {pages} {085117}
  (\bibinfo {year} {2006})}\BibitemShut {NoStop}%
\bibitem [{\citenamefont {Shastry}(2009)}]{Shastry:2009}%
  \BibitemOpen
  \bibfield  {author} {\bibinfo {author} {\bibfnamefont {B.~Sriram}\
  \bibnamefont {Shastry}},\ }\bibfield  {title} {\enquote {\bibinfo {title}
  {Electrothermal transport coefficients at finite frequencies},}\ }\href
  {http://stacks.iop.org/0034-4885/72/i=1/a=016501} {\bibfield  {journal}
  {\bibinfo  {journal} {Reports on Progress in Physics}\ }\textbf {\bibinfo
  {volume} {72}},\ \bibinfo {pages} {016501} (\bibinfo {year}
  {2009})}\BibitemShut {NoStop}%
\bibitem [{\citenamefont {Chakraborty}\ \emph {et~al.}(2010)\citenamefont
  {Chakraborty}, \citenamefont {Galanakis},\ and\ \citenamefont
  {Phillips}}]{Chakraborty:2010}%
  \BibitemOpen
  \bibfield  {author} {\bibinfo {author} {\bibfnamefont {Shiladitya}\
  \bibnamefont {Chakraborty}}, \bibinfo {author} {\bibfnamefont {Dimitrios}\
  \bibnamefont {Galanakis}}, \ and\ \bibinfo {author} {\bibfnamefont {Philip}\
  \bibnamefont {Phillips}},\ }\bibfield  {title} {\enquote {\bibinfo {title}
  {Emergence of particle-hole symmetry near optimal doping in high-temperature
  copper oxide superconductors},}\ }\href {\doibase 10.1103/PhysRevB.82.214503}
  {\bibfield  {journal} {\bibinfo  {journal} {Phys. Rev. B}\ }\textbf {\bibinfo
  {volume} {82}},\ \bibinfo {pages} {214503} (\bibinfo {year}
  {2010})}\BibitemShut {NoStop}%
\bibitem [{\citenamefont {Xu}\ \emph {et~al.}(2011)\citenamefont {Xu},
  \citenamefont {Weber},\ and\ \citenamefont {Kotliar}}]{Xu:2011}%
  \BibitemOpen
  \bibfield  {author} {\bibinfo {author} {\bibfnamefont {Wenhu}\ \bibnamefont
  {Xu}}, \bibinfo {author} {\bibfnamefont {C\'edric}\ \bibnamefont {Weber}}, \
  and\ \bibinfo {author} {\bibfnamefont {Gabriel}\ \bibnamefont {Kotliar}},\
  }\bibfield  {title} {\enquote {\bibinfo {title} {High-frequency
  thermoelectric response in correlated electronic systems},}\ }\href {\doibase
  10.1103/PhysRevB.84.035114} {\bibfield  {journal} {\bibinfo  {journal} {Phys.
  Rev. B}\ }\textbf {\bibinfo {volume} {84}},\ \bibinfo {pages} {035114}
  (\bibinfo {year} {2011})}\BibitemShut {NoStop}%
\bibitem [{\citenamefont {Shastry}\ \emph {et~al.}(1993)\citenamefont
  {Shastry}, \citenamefont {Shraiman},\ and\ \citenamefont
  {Singh}}]{Shastry:1993}%
  \BibitemOpen
  \bibfield  {author} {\bibinfo {author} {\bibfnamefont {B.~Sriram}\
  \bibnamefont {Shastry}}, \bibinfo {author} {\bibfnamefont {Boris~I.}\
  \bibnamefont {Shraiman}}, \ and\ \bibinfo {author} {\bibfnamefont {Rajiv
  R.~P.}\ \bibnamefont {Singh}},\ }\bibfield  {title} {\enquote {\bibinfo
  {title} {Faraday rotation and the hall constant in strongly correlated fermi
  systems},}\ }\href {\doibase 10.1103/PhysRevLett.70.2004} {\bibfield
  {journal} {\bibinfo  {journal} {Phys. Rev. Lett.}\ }\textbf {\bibinfo
  {volume} {70}},\ \bibinfo {pages} {2004--2007} (\bibinfo {year}
  {1993})}\BibitemShut {NoStop}%
\bibitem [{\citenamefont {Assaad}\ and\ \citenamefont
  {Imada}(1995)}]{Assaad:1995}%
  \BibitemOpen
  \bibfield  {author} {\bibinfo {author} {\bibfnamefont {F.~F.}\ \bibnamefont
  {Assaad}}\ and\ \bibinfo {author} {\bibfnamefont {M.}~\bibnamefont {Imada}},\
  }\bibfield  {title} {\enquote {\bibinfo {title} {Hall coefficient for the
  two-dimensional hubbard model},}\ }\href {\doibase
  10.1103/PhysRevLett.74.3868} {\bibfield  {journal} {\bibinfo  {journal}
  {Phys. Rev. Lett.}\ }\textbf {\bibinfo {volume} {74}},\ \bibinfo {pages}
  {3868--3871} (\bibinfo {year} {1995})}\BibitemShut {NoStop}%
\bibitem [{\citenamefont {Kumar}\ and\ \citenamefont
  {Shastry}(2003)}]{Kumar:2003}%
  \BibitemOpen
  \bibfield  {author} {\bibinfo {author} {\bibfnamefont {Brijesh}\ \bibnamefont
  {Kumar}}\ and\ \bibinfo {author} {\bibfnamefont {B.~S.}\ \bibnamefont
  {Shastry}},\ }\bibfield  {title} {\enquote {\bibinfo {title}
  {Superconductivity in ${\mathrm{coo}}_{2}$ layers and the resonating valence
  bond mean-field theory of the triangular lattice $t-j$ model},}\ }\href
  {\doibase 10.1103/PhysRevB.68.104508} {\bibfield  {journal} {\bibinfo
  {journal} {Phys. Rev. B}\ }\textbf {\bibinfo {volume} {68}},\ \bibinfo
  {pages} {104508} (\bibinfo {year} {2003})}\BibitemShut {NoStop}%
\bibitem [{\citenamefont {Kumar}\ and\ \citenamefont
  {Shastry}(2004)}]{Kumar:2004}%
  \BibitemOpen
  \bibfield  {author} {\bibinfo {author} {\bibfnamefont {Brijesh}\ \bibnamefont
  {Kumar}}\ and\ \bibinfo {author} {\bibfnamefont {B.~S.}\ \bibnamefont
  {Shastry}},\ }\bibfield  {title} {\enquote {\bibinfo {title} {Erratum:
  Superconductivity in ${\mathrm{coo}}_{2}$ layers and the resonating valence
  bond mean-field theory of the triangular lattice $t-j$ model [phys. rev. b
  68, 104508 (2003)]},}\ }\href {\doibase 10.1103/PhysRevB.69.059901}
  {\bibfield  {journal} {\bibinfo  {journal} {Phys. Rev. B}\ }\textbf {\bibinfo
  {volume} {69}},\ \bibinfo {pages} {059901} (\bibinfo {year}
  {2004})}\BibitemShut {NoStop}%
\bibitem [{\citenamefont {Haerter}\ and\ \citenamefont
  {Shastry}(2008)}]{Haerter:2008}%
  \BibitemOpen
  \bibfield  {author} {\bibinfo {author} {\bibfnamefont {Jan~O.}\ \bibnamefont
  {Haerter}}\ and\ \bibinfo {author} {\bibfnamefont {B.~Sriram}\ \bibnamefont
  {Shastry}},\ }\bibfield  {title} {\enquote {\bibinfo {title} {Hall number,
  optical sum rule and carrier density for the
  $t\text{-}{t}^{\ensuremath{'}}\text{-}j$ model},}\ }\href {\doibase
  10.1103/PhysRevB.77.045127} {\bibfield  {journal} {\bibinfo  {journal} {Phys.
  Rev. B}\ }\textbf {\bibinfo {volume} {77}},\ \bibinfo {pages} {045127}
  (\bibinfo {year} {2008})}\BibitemShut {NoStop}%
\bibitem [{\citenamefont {Xu}\ \emph {et~al.}(2013)\citenamefont {Xu},
  \citenamefont {Haule},\ and\ \citenamefont {Kotliar}}]{Xu:2013}%
  \BibitemOpen
  \bibfield  {author} {\bibinfo {author} {\bibfnamefont {Wenhu}\ \bibnamefont
  {Xu}}, \bibinfo {author} {\bibfnamefont {Kristjan}\ \bibnamefont {Haule}}, \
  and\ \bibinfo {author} {\bibfnamefont {Gabriel}\ \bibnamefont {Kotliar}},\
  }\bibfield  {title} {\enquote {\bibinfo {title} {Hidden fermi liquid,
  scattering rate saturation, and nernst effect: A dynamical mean-field theory
  perspective},}\ }\href {\doibase 10.1103/PhysRevLett.111.036401} {\bibfield
  {journal} {\bibinfo  {journal} {Phys. Rev. Lett.}\ }\textbf {\bibinfo
  {volume} {111}},\ \bibinfo {pages} {036401} (\bibinfo {year}
  {2013})}\BibitemShut {NoStop}%
\bibitem [{Note1()}]{Note1}%
  \BibitemOpen
  \bibinfo {note} {However, see the recent work \protect \cite
  {OtsukiShinaokaMaxEnt:2017}}\BibitemShut {NoStop}%
\bibitem [{\citenamefont {Reymbaut}\ \emph {et~al.}(2015)\citenamefont
  {Reymbaut}, \citenamefont {Bergeron},\ and\ \citenamefont
  {Tremblay}}]{Reymbaut:2015_MaxEnt}%
  \BibitemOpen
  \bibfield  {author} {\bibinfo {author} {\bibfnamefont {A.}~\bibnamefont
  {Reymbaut}}, \bibinfo {author} {\bibfnamefont {D.}~\bibnamefont {Bergeron}},
  \ and\ \bibinfo {author} {\bibfnamefont {A.-M.~S.}\ \bibnamefont
  {Tremblay}},\ }\bibfield  {title} {\enquote {\bibinfo {title} {Maximum
  entropy analytic continuation for spectral functions with nonpositive
  spectral weight},}\ }\href {\doibase 10.1103/PhysRevB.92.060509} {\bibfield
  {journal} {\bibinfo  {journal} {Phys. Rev. B}\ }\textbf {\bibinfo {volume}
  {92}},\ \bibinfo {pages} {060509(R)} (\bibinfo {year} {2015})}\BibitemShut
  {NoStop}%
\bibitem [{\citenamefont {Reymbaut}\ \emph {et~al.}(2016)\citenamefont
  {Reymbaut}, \citenamefont {Charlebois}, \citenamefont {Asiani}, \citenamefont
  {Fratino}, \citenamefont {S\'emon}, \citenamefont {Sordi},\ and\
  \citenamefont {Tremblay}}]{Reymbaut:2016_Cuprates_1}%
  \BibitemOpen
  \bibfield  {author} {\bibinfo {author} {\bibfnamefont {A.}~\bibnamefont
  {Reymbaut}}, \bibinfo {author} {\bibfnamefont {M.}~\bibnamefont
  {Charlebois}}, \bibinfo {author} {\bibfnamefont {M.~Fellous}\ \bibnamefont
  {Asiani}}, \bibinfo {author} {\bibfnamefont {L.}~\bibnamefont {Fratino}},
  \bibinfo {author} {\bibfnamefont {P.}~\bibnamefont {S\'emon}}, \bibinfo
  {author} {\bibfnamefont {G.}~\bibnamefont {Sordi}}, \ and\ \bibinfo {author}
  {\bibfnamefont {A.-M.~S.}\ \bibnamefont {Tremblay}},\ }\bibfield  {title}
  {\enquote {\bibinfo {title} {Antagonistic effects of nearest-neighbor
  repulsion on the superconducting pairing dynamics in the doped mott insulator
  regime},}\ }\href {\doibase 10.1103/PhysRevB.94.155146} {\bibfield  {journal}
  {\bibinfo  {journal} {Phys. Rev. B}\ }\textbf {\bibinfo {volume} {94}},\
  \bibinfo {pages} {155146} (\bibinfo {year} {2016})}\BibitemShut {NoStop}%
\bibitem [{\citenamefont {Gagnon}(2016)}]{GagnonMSc:2016}%
  \BibitemOpen
  \bibfield  {author} {\bibinfo {author} {\bibfnamefont {A.-M.}\ \bibnamefont
  {Gagnon}},\ }\emph {\bibinfo {title} {Une méthode alternative pour obtenir
  le pouvoir thermoélectrique à température finie}},\ \href
  {http://www.physique.usherbrooke.ca/pages/sites/default/files/Gagnon_AM_Msc.pdf}
  {\bibinfo {type} {M.sc.}},\ \bibinfo  {school} {Université de Sherbrooke}
  (\bibinfo {year} {2016})\BibitemShut {NoStop}%
\bibitem [{\citenamefont {Georges}\ \emph {et~al.}(1996)\citenamefont
  {Georges}, \citenamefont {Kotliar}, \citenamefont {Krauth},\ and\
  \citenamefont {Rozenberg}}]{Georges:1996}%
  \BibitemOpen
  \bibfield  {author} {\bibinfo {author} {\bibfnamefont {A.}~\bibnamefont
  {Georges}}, \bibinfo {author} {\bibfnamefont {G.}~\bibnamefont {Kotliar}},
  \bibinfo {author} {\bibfnamefont {W.}~\bibnamefont {Krauth}}, \ and\ \bibinfo
  {author} {\bibfnamefont {M.~J.}\ \bibnamefont {Rozenberg}},\ }\bibfield
  {title} {\enquote {\bibinfo {title} {Dynamical mean-field theory of strongly
  correlated fermion systems and the limit of infinite dimensions},}\ }\href
  {http://journals.aps.org/rmp/abstract/10.1103/RevModPhys.68.13} {\bibfield
  {journal} {\bibinfo  {journal} {Rev. Mod. Phys.}\ }\textbf {\bibinfo {volume}
  {68}},\ \bibinfo {pages} {13--25} (\bibinfo {year} {1996})}\BibitemShut
  {NoStop}%
\bibitem [{\citenamefont {Gull}\ \emph {et~al.}(2011)\citenamefont {Gull},
  \citenamefont {Millis}, \citenamefont {Lichtenstein}, \citenamefont
  {Rubtsov}, \citenamefont {Troyer},\ and\ \citenamefont {Werner}}]{Gull:2011}%
  \BibitemOpen
  \bibfield  {author} {\bibinfo {author} {\bibfnamefont {Emanuel}\ \bibnamefont
  {Gull}}, \bibinfo {author} {\bibfnamefont {Andrew~J.}\ \bibnamefont
  {Millis}}, \bibinfo {author} {\bibfnamefont {Alexander~I.}\ \bibnamefont
  {Lichtenstein}}, \bibinfo {author} {\bibfnamefont {Alexey~N.}\ \bibnamefont
  {Rubtsov}}, \bibinfo {author} {\bibfnamefont {Matthias}\ \bibnamefont
  {Troyer}}, \ and\ \bibinfo {author} {\bibfnamefont {Philipp}\ \bibnamefont
  {Werner}},\ }\bibfield  {title} {\enquote {\bibinfo {title} {Continuous-time
  {Monte Carlo} methods for quantum impurity models},}\ }\href {\doibase
  10.1103/RevModPhys.83.349} {\bibfield  {journal} {\bibinfo  {journal} {Rev.
  Mod. Phys.}\ }\textbf {\bibinfo {volume} {83}},\ \bibinfo {pages} {349--404}
  (\bibinfo {year} {2011})}\BibitemShut {NoStop}%
\bibitem [{\citenamefont {Randeria}\ \emph {et~al.}(1992)\citenamefont
  {Randeria}, \citenamefont {Trivedi}, \citenamefont {Moreo},\ and\
  \citenamefont {Scalettar}}]{Randeria:1992}%
  \BibitemOpen
  \bibfield  {author} {\bibinfo {author} {\bibfnamefont {Mohit}\ \bibnamefont
  {Randeria}}, \bibinfo {author} {\bibfnamefont {Nandini}\ \bibnamefont
  {Trivedi}}, \bibinfo {author} {\bibfnamefont {Adriana}\ \bibnamefont
  {Moreo}}, \ and\ \bibinfo {author} {\bibfnamefont {Richard~T.}\ \bibnamefont
  {Scalettar}},\ }\bibfield  {title} {\enquote {\bibinfo {title} {Pairing and
  spin gap in the normal state of short coherence length superconductors},}\
  }\href {\doibase 10.1103/PhysRevLett.69.2001} {\bibfield  {journal} {\bibinfo
   {journal} {Phys. Rev. Lett.}\ }\textbf {\bibinfo {volume} {69}},\ \bibinfo
  {pages} {2001--2004} (\bibinfo {year} {1992})}\BibitemShut {NoStop}%
\bibitem [{Note2()}]{Note2}%
  \BibitemOpen
  \bibinfo {note} {See supplemental material at [] for the choice of convention
  for the Onsager coefficients (identical to that of Ref.~\protect \cite
  {Mahan:2000}). Explanations behind the low-temperature approximation
  Eq.~\protect \eqref {Eq_L_Randeria}~\protect \cite {Randeria:1992} are given.
  We also present results at $U=14$, $T = 1/10$ and $T=1$ including the
  response functions' spectral weights along with a more thorough comparison
  with Ref.~\protect \cite {Xu:2011} and the approximation Eq.~\protect \eqref
  {Eq_L_Randeria}. It is also shown how the results are affected by the choice
  of $\lambda $ in Eqs.~\protect \eqref {Eq_mixed_operator-agd} and \protect
  \eqref {Eq_mixed_operator_b}. We end with new expressions~\protect \cite
  {GagnonMSc:2016} for the bubble part of the uniform susceptibilities that are
  convergent upon summation over internal Matsubara frequencies~\protect \cite
  {Bergeron:2011}. The convergence of the corresponding expressions given in
  Ref.~\cite {Paul:2003} is discussed. In addition, we include examples of data
  files used for obtaining the first figure of this paper with
  OmegaMaxEnt~\protect \cite {Bergeron:2015}. The supplemental material also
  contains a set of data files enabling the user to reproduce the T=1 and
  lambda=0.3 results of the parent paper using the OmegaMaxEnt code~\protect
  \cite {Bergeron:2015}.}\BibitemShut {Stop}%
\bibitem [{\citenamefont {Spielman}\ \emph {et~al.}(1994)\citenamefont
  {Spielman}, \citenamefont {Parks}, \citenamefont {Orenstein}, \citenamefont
  {Nemeth}, \citenamefont {Ludwig}, \citenamefont {Clarke}, \citenamefont
  {Merchant},\ and\ \citenamefont {Lew}}]{Spielman:1994}%
  \BibitemOpen
  \bibfield  {author} {\bibinfo {author} {\bibfnamefont {S.}~\bibnamefont
  {Spielman}}, \bibinfo {author} {\bibfnamefont {Beth}\ \bibnamefont {Parks}},
  \bibinfo {author} {\bibfnamefont {J.}~\bibnamefont {Orenstein}}, \bibinfo
  {author} {\bibfnamefont {D.~T.}\ \bibnamefont {Nemeth}}, \bibinfo {author}
  {\bibfnamefont {Frank}\ \bibnamefont {Ludwig}}, \bibinfo {author}
  {\bibfnamefont {John}\ \bibnamefont {Clarke}}, \bibinfo {author}
  {\bibfnamefont {Paul}\ \bibnamefont {Merchant}}, \ and\ \bibinfo {author}
  {\bibfnamefont {D.~J.}\ \bibnamefont {Lew}},\ }\bibfield  {title} {\enquote
  {\bibinfo {title} {Observation of the quasiparticle hall effect in
  superconducting
  $\mathrm{Y}{\mathrm{ba}}_{2}{\mathrm{cu}}_{3}{\mathrm{o}}_{7-\ensuremath{\delta}}$},}\
  }\href {\doibase 10.1103/PhysRevLett.73.1537} {\bibfield  {journal} {\bibinfo
   {journal} {Phys. Rev. Lett.}\ }\textbf {\bibinfo {volume} {73}},\ \bibinfo
  {pages} {1537--1540} (\bibinfo {year} {1994})}\BibitemShut {NoStop}%
\bibitem [{\citenamefont {Drew}\ and\ \citenamefont
  {Coleman}(1997)}]{Drew_Coleman:1997}%
  \BibitemOpen
  \bibfield  {author} {\bibinfo {author} {\bibfnamefont {H.~D.}\ \bibnamefont
  {Drew}}\ and\ \bibinfo {author} {\bibfnamefont {P.}~\bibnamefont {Coleman}},\
  }\bibfield  {title} {\enquote {\bibinfo {title} {Sum rule for the optical
  hall angle},}\ }\href {\doibase 10.1103/PhysRevLett.78.1572} {\bibfield
  {journal} {\bibinfo  {journal} {Phys. Rev. Lett.}\ }\textbf {\bibinfo
  {volume} {78}},\ \bibinfo {pages} {1572--1575} (\bibinfo {year}
  {1997})}\BibitemShut {NoStop}%
\bibitem [{\citenamefont {Arsenault}\ and\ \citenamefont
  {Tremblay}(2013)}]{LFA:2013}%
  \BibitemOpen
  \bibfield  {author} {\bibinfo {author} {\bibfnamefont
  {Louis-Fran\mbox{\c{c}}ois}\ \bibnamefont {Arsenault}}\ and\ \bibinfo
  {author} {\bibfnamefont {A.-M.~S.}\ \bibnamefont {Tremblay}},\ }\bibfield
  {title} {\enquote {\bibinfo {title} {Transport functions for hypercubic and
  bethe lattices},}\ }\href {\doibase 10.1103/PhysRevB.88.205109} {\bibfield
  {journal} {\bibinfo  {journal} {Phys. Rev. B}\ }\textbf {\bibinfo {volume}
  {88}},\ \bibinfo {pages} {205109} (\bibinfo {year} {2013})}\BibitemShut
  {NoStop}%
\bibitem [{\citenamefont {Bergeron}\ and\ \citenamefont
  {Tremblay}(2016)}]{Bergeron:2015}%
  \BibitemOpen
  \bibfield  {author} {\bibinfo {author} {\bibfnamefont {Dominic}\ \bibnamefont
  {Bergeron}}\ and\ \bibinfo {author} {\bibfnamefont {A.-M.~S.}\ \bibnamefont
  {Tremblay}},\ }\bibfield  {title} {\enquote {\bibinfo {title} {Algorithms for
  optimized maximum entropy and diagnostic tools for analytic continuation},}\
  }\href {\doibase 10.1103/PhysRevE.94.023303} {\bibfield  {journal} {\bibinfo
  {journal} {Phys. Rev. E}\ }\textbf {\bibinfo {volume} {94}},\ \bibinfo
  {pages} {023303} (\bibinfo {year} {2016})}\BibitemShut {NoStop}%
\bibitem [{\citenamefont {Mahan}(2000)}]{Mahan:2000}%
  \BibitemOpen
  \bibfield  {author} {\bibinfo {author} {\bibfnamefont {G.~D.}\ \bibnamefont
  {Mahan}},\ }\href@noop {} {\emph {\bibinfo {title} {Many-Particle Physics,
  3rd edition, Section 6.4.4}}}\ (\bibinfo {year} {2000})\BibitemShut {NoStop}%
\bibitem [{\citenamefont {{Chen}}\ \emph {et~al.}(2016)\citenamefont {{Chen}},
  \citenamefont {{LeBlanc}},\ and\ \citenamefont {{Gull}}}]{Chen:2016}%
  \BibitemOpen
  \bibfield  {author} {\bibinfo {author} {\bibfnamefont {X.}~\bibnamefont
  {{Chen}}}, \bibinfo {author} {\bibfnamefont {J.~P.~F.}\ \bibnamefont
  {{LeBlanc}}}, \ and\ \bibinfo {author} {\bibfnamefont {E.}~\bibnamefont
  {{Gull}}},\ }\bibfield  {title} {\enquote {\bibinfo {title} {{Knight shifts,
  nuclear spin-relaxation rates, and spin echo decay times in the pseudogap
  regime of the cuprates: Simulation and relation to experiment}},}\ }\href
  {http://adsabs.harvard.edu/abs/2016arXiv160705655C} {\bibfield  {journal}
  {\bibinfo  {journal} {ArXiv e-prints}\ } (\bibinfo {year} {2016})},\ \Eprint
  {http://arxiv.org/abs/1607.05655} {arXiv:1607.05655 [cond-mat.str-el]}
  \BibitemShut {NoStop}%
\bibitem [{Note3()}]{Note3}%
  \BibitemOpen
  \bibinfo {note} {Notice that this approximation is not valid in the
  renormalized classical regime since, in this regime, the convergence of the
  integral in Eq.~\protect \eqref {Eq_L_Randeria} is controlled by the
  susceptibility instead of the $\protect \qopname \relax o{sinh}$
  factor.}\BibitemShut {Stop}%
\bibitem [{\citenamefont {{Otsuki}}\ \emph {et~al.}(2017)\citenamefont
  {{Otsuki}}, \citenamefont {{Ohzeki}}, \citenamefont {{Shinaoka}},\ and\
  \citenamefont {{Yoshimi}}}]{OtsukiShinaokaMaxEnt:2017}%
  \BibitemOpen
  \bibfield  {author} {\bibinfo {author} {\bibfnamefont {J.}~\bibnamefont
  {{Otsuki}}}, \bibinfo {author} {\bibfnamefont {M.}~\bibnamefont {{Ohzeki}}},
  \bibinfo {author} {\bibfnamefont {H.}~\bibnamefont {{Shinaoka}}}, \ and\
  \bibinfo {author} {\bibfnamefont {K.}~\bibnamefont {{Yoshimi}}},\ }\bibfield
  {title} {\enquote {\bibinfo {title} {{Sparse modeling approach to analytical
  continuation of imaginary-time quantum Monte Carlo data}},}\ }\href@noop {}
  {\bibfield  {journal} {\bibinfo  {journal} {ArXiv e-prints}\ } (\bibinfo
  {year} {2017})},\ \Eprint {http://arxiv.org/abs/1702.03056} {arXiv:1702.03056
  [cond-mat.str-el]} \BibitemShut {NoStop}%
\bibitem [{\citenamefont {Bergeron}\ \emph {et~al.}(2011)\citenamefont
  {Bergeron}, \citenamefont {Hankevych}, \citenamefont {Kyung},\ and\
  \citenamefont {Tremblay}}]{Bergeron:2011}%
  \BibitemOpen
  \bibfield  {author} {\bibinfo {author} {\bibfnamefont {Dominic}\ \bibnamefont
  {Bergeron}}, \bibinfo {author} {\bibfnamefont {Vasyl}\ \bibnamefont
  {Hankevych}}, \bibinfo {author} {\bibfnamefont {Bumsoo}\ \bibnamefont
  {Kyung}}, \ and\ \bibinfo {author} {\bibfnamefont {A.-M.~S.}\ \bibnamefont
  {Tremblay}},\ }\bibfield  {title} {\enquote {\bibinfo {title} {Optical and dc
  conductivity of the two-dimensional hubbard model in the pseudogap regime and
  across the antiferromagnetic quantum critical point including vertex
  corrections},}\ }\href {\doibase 10.1103/PhysRevB.84.085128} {\bibfield
  {journal} {\bibinfo  {journal} {Phys. Rev. B}\ }\textbf {\bibinfo {volume}
  {84}},\ \bibinfo {pages} {085128} (\bibinfo {year} {2011})}\BibitemShut
  {NoStop}%
\bibitem [{\citenamefont {Paul}\ and\ \citenamefont
  {Kotliar}(2003)}]{Paul:2003}%
  \BibitemOpen
  \bibfield  {author} {\bibinfo {author} {\bibfnamefont {Indranil}\
  \bibnamefont {Paul}}\ and\ \bibinfo {author} {\bibfnamefont {Gabriel}\
  \bibnamefont {Kotliar}},\ }\bibfield  {title} {\enquote {\bibinfo {title}
  {Thermal transport for many-body tight-binding models},}\ }\href {\doibase
  10.1103/PhysRevB.67.115131} {\bibfield  {journal} {\bibinfo  {journal} {Phys.
  Rev. B}\ }\textbf {\bibinfo {volume} {67}},\ \bibinfo {pages} {115131}
  (\bibinfo {year} {2003})}\BibitemShut {NoStop}%
\end{thebibliography}

\end{document}